\documentclass[journal]{IEEEtran}
\usepackage{cite}
\usepackage{amsmath,amssymb,amsfonts}
\usepackage[ruled,vlined]{algorithm2e}
\usepackage{graphicx}
\usepackage{textcomp}
\usepackage{xcolor}
\usepackage{multirow}
\usepackage{booktabs}
\usepackage{subcaption}
\usepackage{hyperref}
\usepackage{todonotes}
\usepackage{color,soul}
\usepackage{xcolor}
\newcommand\blfootnote[1]{%
  \begingroup
  \renewcommand\thefootnote{}\footnote{#1}%
  \addtocounter{footnote}{-1}%
  \endgroup
}

\hypersetup{
     colorlinks = true,
     linkcolor = blue,
     anchorcolor = blue,
     citecolor = green,
     filecolor = magenta,
     urlcolor = cyan
     }

\ifCLASSINFOpdf
\else
\fi
\begin{document}
\title{$N^2C:$ Neural Network Controller Design Using Behavioral Cloning}

\author{Shoaib~Azam,~\IEEEmembership{Student Member,~IEEE,}
         Farzeen~Munir,~\IEEEmembership{Student Member,~IEEE,}
         Muhammad~Aasim~Rafique,~\IEEEmembership{Member,~IEEE,}
		 Ahmad~Muqeem~Sheri,~\IEEEmembership{Member,~IEEE,}
		 Muhammad Ishfaq Hussain,~\IEEEmembership{Student Member,~IEEE,}
        and Moongu~Jeon,~\IEEEmembership{Senior Member,~IEEE,}% <-this % stops a space
\thanks{Shoaib~Azam, Farzeen~Munir, Muhammad Ishfaq Hussain, and Moongu~Jeon are 
with the School of Electrical Engineering and Computer Science, 
Gwangju Institute of Science and Technology,
Gwangju, South Korea. 
\newline Muhammad Aasim Rafique and Ahmed Muqeem Sheri are  with National University of Sciences and Technology (NUST), Islamabad, Pakistan. 
% Ahmad Muqeem Sheri is with Department of Computer Software Engineering, National University of Sciences and Technology (NUST), Islamabad, Pakistan.
e-mail: (shoaibazam@gist.ac.kr; farzeen.munir@gist.ac.kr; ishfaqhussain@gist.ac.kr; aasim.rafique@seecs.edu.pk; muqeem@mcs.edu.pk; mgjeon@gist.ac.kr}}
\markboth{Journal of \LaTeX\ Class Files,~Vol.~14, No.~8, August~2015}%
{Shell \MakeLowercase{\textit{et al.}}: Bare Demo of IEEEtran.cls for IEEE Journals}



\maketitle

\begin{abstract}
Modern vehicles communicate data to and from sensors, actuators, and electronic control units (ECUs) using Controller Area Network (CAN) bus, which operates on differential signaling. An autonomous ECU responsible for the execution of decision commands to an autonomous vehicle is developed by assimilating the information from the CAN bus. The conventional way of parsing the decision commands is motion planning, which uses a path tracking algorithm to evaluate the decision commands. This study focuses on designing a robust controller using behavioral cloning and motion planning of autonomous vehicle using a deep learning framework. In the first part of this study, we explore the pipeline of parsing decision commands from the path tracking algorithm to the controller and proposed a neural network-based controller ($N^2C$) using behavioral cloning. The proposed network predicts throttle, brake, and torque when trained with the manual driving data acquired from the CAN bus. The efficacy of the proposed method is demonstrated by comparing the accuracy with the Proportional-Derivative-Integral (PID) controller in conjunction with the path tracking algorithm (pure pursuit and model predictive control based path follower). The second part of this study complements $N^2C$, in which an end-to-end neural network for predicting the speed and steering angle is proposed with image data as an input. The performance of the proposed frameworks are evaluated in real-time and on the Udacity dataset, showing better metric scores in the former and reliable prediction in the later case when compared with the state-of-the-art methods.

\end{abstract}


% \fi
\IEEEpeerreviewmaketitle

\section{Introduction}
\IEEEPARstart{A}utonomous driving is one of the most computation extensive areas of the automobile industry and demands the development of new techniques utilizing mature sensors suite. The automation level in an autonomous vehicle is introduced by the Society of Automation Engineering (SAE) where, level-0 is manual control, and level-5 corresponds to fully autonomous mode\footnote{https://www.sae.org/news/2019/01/sae-updates-j3016-automated-driving-graphic}. Besides, an increase in the automation level, safety of the autonomous vehicle in a dynamic traffic environment is a primary concern of many stakeholders, including academicians, industrialists, researchers, and policymakers. Engineers and researchers are striving to enhance the safety level. The safety measures as standardized by SOTIF-ISO/PAS-21448 (Safety of the intended functionality) incorporate both the hardware and software aspects of the autonomous vehicle to provide complete situational awareness without system failure\footnote{https://www.daimler.com/innovation/case/autonomous/safety-first-for-automated-driving-2.htm}. 

% owadays autonomous driving is one of the main focused areas of the automobile industry, and many stakeholders, including academics, industrialists, and researchers are striving their best to enhance the automation level. The Society of Automation Engineering (SAE) has devised six standards of automation in which level-0 is manual control, and level-5 corresponds to fully autonomous mode \cite{Sae}.The successful increase in automation level demands the development in both hardware and software of the vehicle. The self-driving vehicle is equipped with sensors like cameras, Lidars, radars, Global Navigation Satellite System (GNSS), and state-of-the-art algorithms to solve the problems of localization, perception, navigation, and control.

\begin{figure}[t]
      \centering
      \includegraphics[width=7cm]{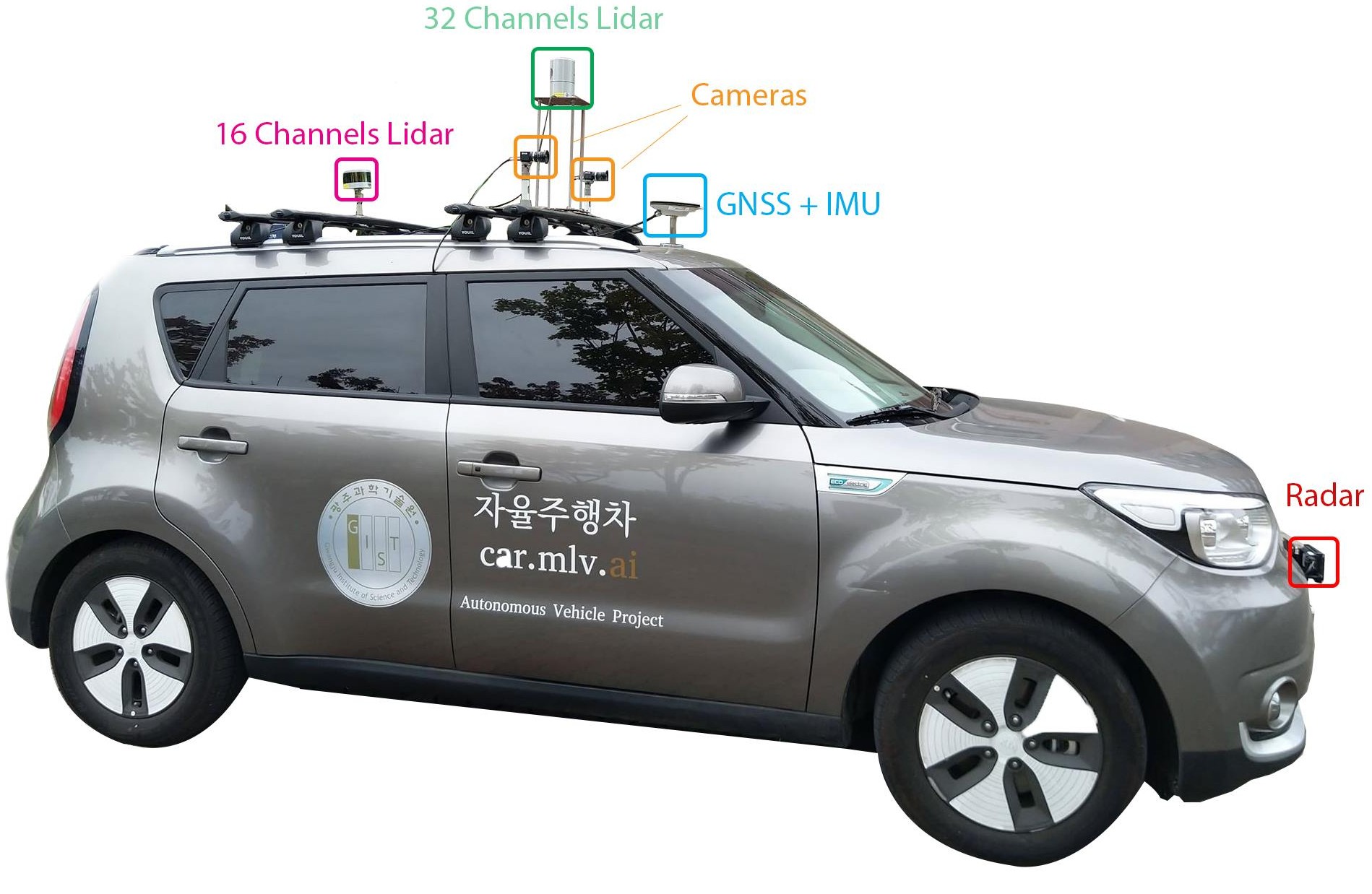}
      \caption{Our autonomous vehicle is equipped with state-of-the-art sensors.}
      \label{car}
\end{figure}
% \IEEEPARstart{N}owadays autonomous driving is one of the main focused area of automobile industry and many stakeholders including academics, industrialist and researchers are striving their best to ameliorate the automation level. The Society of Automation Engineering (SAE) has devised six standards of automation in which level-0 is manual control and level-5 corresponds to fully autonomous mode \cite{1}. The successful increase in automation level demands the development in both hardware and software directions, and to meet this condition, the self-driving vehicle is equipped with many sensors like cameras, Lidars, radars, Global Navigation Satellite System (GNSS) along with state-of-the-art algorithms for the solving the problems of localization, perception, navigation and control. 
\par 
An accurately assessed response (planning) and a precise execution (control) of the response is a desiderata of safety in autonomous vehicles. Besides external sensors, modern vehicles are equipped with many in-built sensors and electronic control units (ECUs) that provide extensive and real-time information about the car for the driver's safety and control. This massive amount of data is available by the Controller-Area-Network (CAN) bus developed by Robert Bosch in 1986 \cite{kiencke1986automotive} that serves as a communication channel between different electronic components of the car. The longitudinal and lateral control of the autonomous vehicle maps the planned desired actions to the vehicle actuators through the CAN bus.

\par 
A complete framework of autonomous vehicle includes localization, perception, and planning modules. Localization enables an autonomous vehicle to localize itself in a dynamic environment. The two essentials entities required for localization are $3$D maps (offline or online) and map-matching algorithms. The localization of the autonomous vehicle is done using Normal Distribution Transform matching \cite{munir2019localization}. 

\begin{figure*}[t]
      \centering
      \includegraphics[width=15cm]{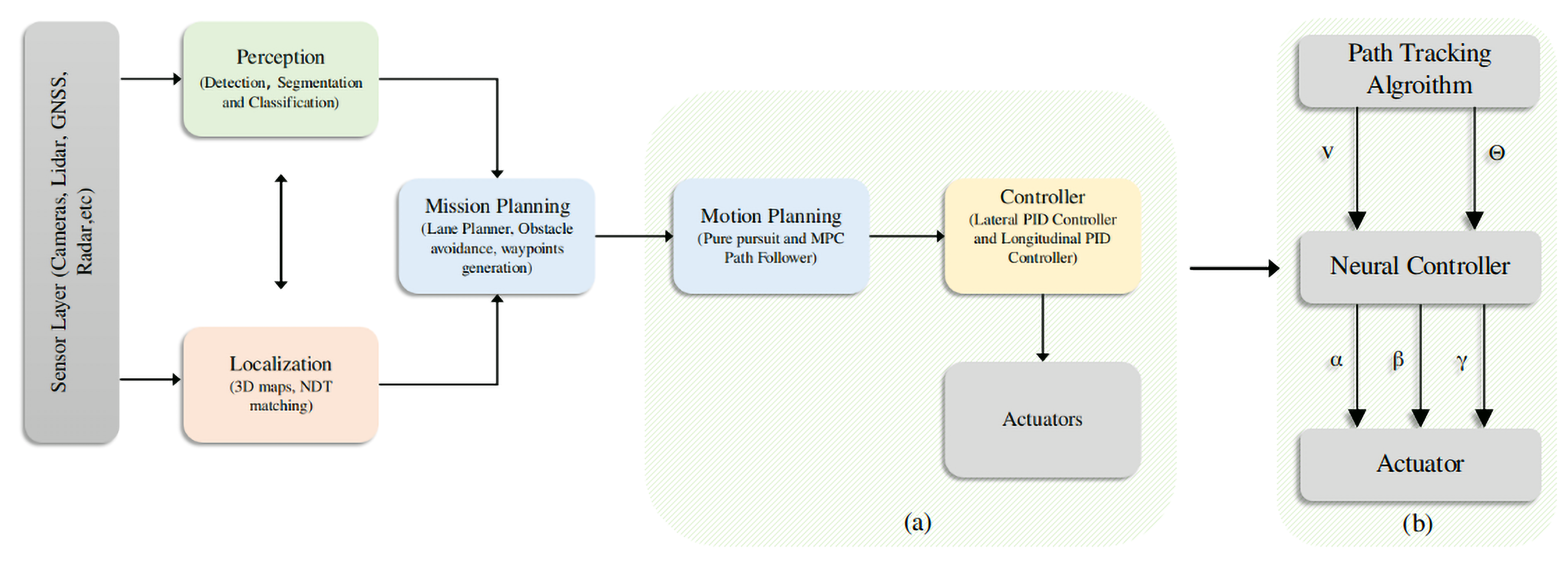}
      \caption{The overall architecture of the autonomous vehicle is shown in this figure.  (a) The conventional pipeline from motion planning to actuators using classical controllers \cite{azam2020dynamic}. (b) The proposed framework for designing a neural network-based controller. $v$ and $\theta$ are linear and angular velocities from the path tracking algorithm. $\alpha$, $\beta$, and $\gamma$ are predicted throttle, brake, and torque from the neural network-based controller.  }
      \label{architecture}
\end{figure*}

The perception module incorporates the classification of objects in the environment through commonly used sensor modalities that are lidar and camera. In the perception module, object detection, classification, and segmentation are done using lidar and image data. Yolo \cite{farhadi2018yolov3} and SSD \cite{liu2016ssd} are used to perform image-based object detection, whereas euclidean clustering is used for lidar based object detection. SegNet is used for segmenting images \cite{badrinarayanan2017segnet}, and PointPillars is used for segmentation and classification using lidar data \cite{lang2019pointpillars}. The perception of an environment provides the necessary information about the environment, which is used by the planning module for motion and mission planning. In mission planning, a lane planner is devised by defining a state machine that tackles the object avoidance by incorporating the information from perception module. The mission planning generates the optimal waypoints used by the motion planning modules, including path tracking algorithms. Finally, the path tracking algorithm provides the desired linear and angular velocities information to the controller.
\par 
The control module plays an important role in predicting the behavior of an autonomous vehicle. There are two main categories of vehicle control i) longitudinal control, and ii) lateral control. The vehicle longitudinal control deals with vehicle speed regulation, and the lateral control is responsible for the sideways movement. In other words, vehicle speed and brake are managed by longitudinal control, whereas the steering angle is controlled by lateral control. The longitudinal and lateral control commands are transmitted to vehicle actuators through the CAN bus in terms of forces that includes throttle, brake, and torque. Generally,  two separate controllers are utilized for controlling the autonomous vehicle. In literature, researchers have used different control strategies that include a Proportional-Integral-Derivative (PID) controller\cite{perez2011cascade}, predictive control mechanisms\cite{cristi1990adaptive}, and model-reference adaptive control methods\cite{borrelli2005mpc}. 
\par 
The control and planning modules are associated in the stack of an autonomous vehicle using a path tracking algorithm. The planning module formulates an optimized path based on poses and velocities using mission and motion planning. A path tracking algorithm computes the linear and angular velocity, which is fed to the controller for longitudinal and lateral movements. In this study, we explore the pipeline from the path-tracking algorithm to the controller, which includes i) path-tracking algorithm ii) longitudinal controller and iii) lateral controller using behavioral cloning. Fig. \ref{architecture}(b) shows a layout of the proposed framework. This study divides the proposed framework into two parts for ease of understanding, where the first part focuses on the design of a neural network-based controller ($N^2C$) using behavioral cloning \cite{torabi2018behavioral} \cite{zhang2018behavioral}. The second part uses image data for the prediction of speed and steering angle and a deep learning surrogates path tracking algorithm. For the neural network-based controller ($N^2C$) using behavioral cloning, the human driving data is collected that includes throttle (drive pedal), brake (brake pedal), torque (steering torque), speed and steering angle through CAN bus along with corresponding image data. 
\par

Recent research has introduced an end-to-end deep neural network to predict the steering angle and speed from images. However, the values of steering angle and speed predicted by the deep neural network cannot be verified with a real-time experiment. For instance, the steering angle predicted by the network is not in the limits of actuator \cite{qiu2019learning}. In the second part of this work, we explore a strategy to verify the predictions of end-to-end deep neural network by predicting the steering angle and speed using images and feeding it to a neural network-based controller ($N^2C$), which in return generates throttle, brake, and torque for the vehicle actuators.

\subsection{Background and Motivation}
To develop a testbed for the autonomous vehicle, we modify an electric KIA Soul EV car. Fig. \ref{car} shows our autonomous vehicle testbed equipped with perception and navigational sensors. In the perception sensors, lidar and cameras have been used for acquiring data for the perception of the environment. The location data is acquired using navigation sensors, which include a Global Navigation Satellite System (GNSS). It is an enclosed device containing the Global positing System (GPS) and Inertial Measurement Unit (IMU). The GPS gives the location with a tolerance of $2cm$ \cite{munir2018autonomous}. Fig. \ref{architecture} shows the architecture of our autonomous vehicle.
% In order to conduct this study, we modify an electric KIA soul EV car to a self-driving vehicle. Fig.\ref{car}shows our self-driving vehicle test-bed equipped with perception and navigational sensors. In the perception sensors, Lidar and cameras have used for the acquiring the perception of the environment. The localization of self-driving is done using the navigation sensors which includes Novatel Global Navigation Satellite System (GNSS). This is single enclosed device containing Global positing System (GPS) and Inertial Measurement Unit (IMU). The GPS gives the location with tolerance of $2cm$. We have used Robot Operating System (ROS) for the acquiring data from sensors and algorithms implementation. Fig.\ref{architecture} show the architecture of our self-driving vehicle.

\par 
The primary motivation of this work is to conduct an empirical study of replacement of the current modules of control architecture by neural network based on behavioral cloning. The experimental evaluation illustrates its advantages in terms of smoothness and robustness, for using $N^2C$ and conventional control methods in juxtaposition. Second, this neural network-based controller model does not require human supervision for manual tuning of parameters as compared to conventional classical control theory. 
\par 
The main contributions are listed as follows:
\begin{enumerate}
    \item A novel controller is designed based on behavioral cloning using a neural network.
    \item The designed neural network controller is utilized with the end-to-end approach of predicting steering angle and speed. 
    \item We have utilized the designed neural network-based controller ($N^2C$) with two different path-tracking algorithms to assess the efficacy of the proposed controller compared to the conventional classical controller.
\end{enumerate}

\par
The rest of the paper is organized as follows: Section II gives related work. Section III focuses on the proposed methodology. The experimentation and results are discussed in Section IV, and section V concludes the paper.

%\IEEEpubidadjcol
 \begin{figure*}[t]
      \centering
      \includegraphics[width=18cm]{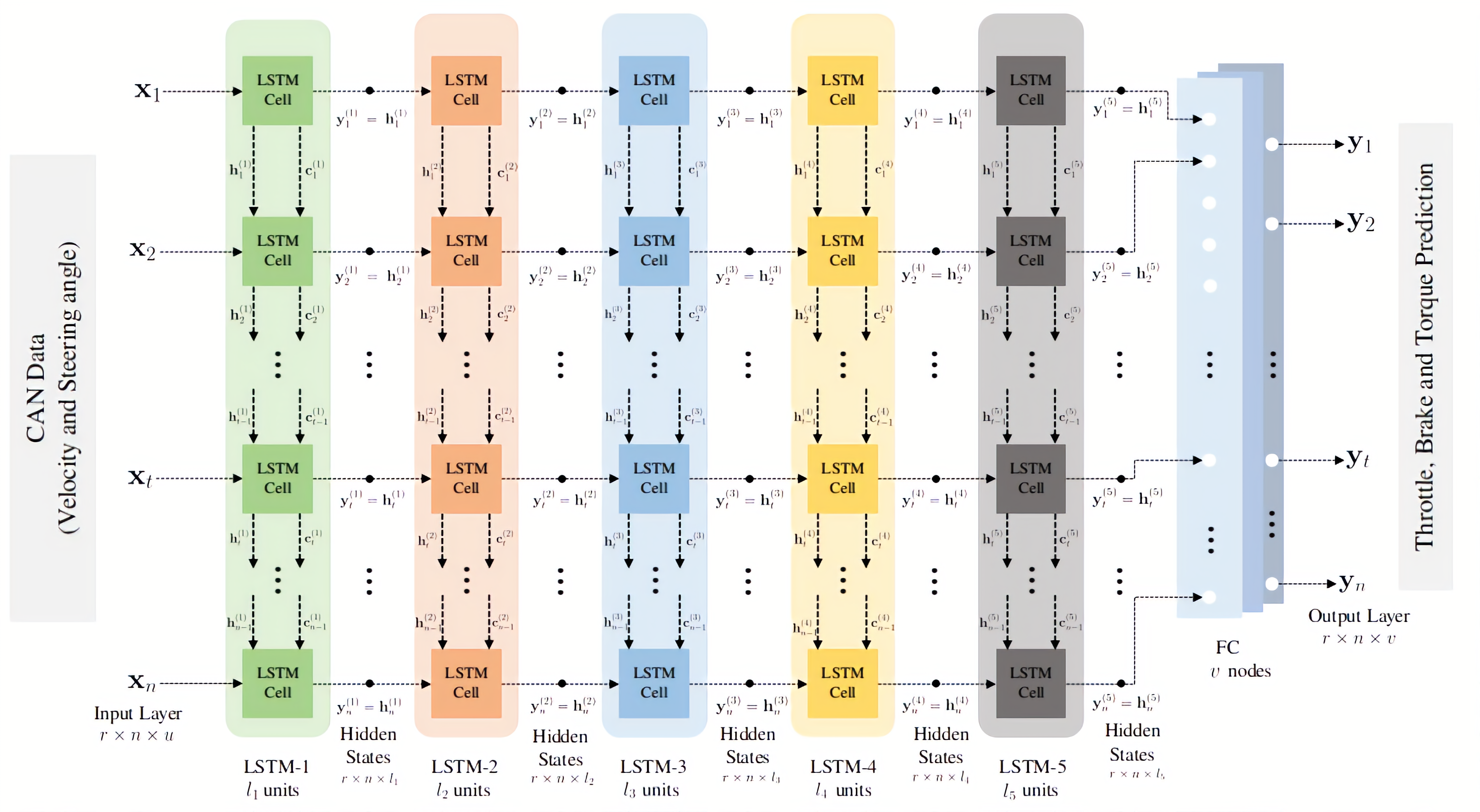}
      \caption{The proposed framework for a controller, $N^2C$ consisting of multi-layer LSTM architecture.}
      \label{framework}
\end{figure*}

\section{Related Work}
\subsection{Control for autonomous vehicle}
The control module of the autonomous vehicle is responsible for communicating with car electronics through CAN bus Fig. \ref{architecture}. It sends appropriate commands of the throttle, brake, and steering torque based on the angle and speed determine by the planning module to ECU \cite {kritayakirana2012autonomous}. In literature, the conventional control modules consist of longitudinal and lateral controllers that enable the autonomous vehicle to follow a planned trajectory \cite{khodayari2010historical}. These longitudinal and lateral controllers are designed using conventional approaches such as PID control \cite{perez2011cascade}, Linear-Quadratic Regulator control \cite{cristi1990adaptive}, Model predictive control \cite{borrelli2005mpc}, fuzzy logic \cite{kodagoda2002fuzzy}, and so on. In recent years, the researchers are experimenting with classical control approaches with intelligent parameters tuning. \cite{wang2019speed} has proposed a fuzzy self-tuning PID controller for an autonomous vehicle. They have adopted a fuzzy logic on the top of the PID controller to address the problem of frequent switching between throttle and brake and also for optimizing the PID parameters online for nonlinear systems. \cite{han2019design} has used the artificial neural network to tune the PID parameters. \cite{wang2020adaptive} has combined the feed-forward PID controller with self-adapting PID to improve the accuracy of the control algorithm.

\par 
Model Predictive Control (MPC) or receding horizon control optimizes the control process by satisfying the constraints induced by the system dynamics model. MPC predicts the change in the future values in an iterative manner, by optimizing the cost function over the receding horizon. \cite{falcone2008mpc} has proposed the MPC based controller for steer and brake paddle control of an autonomous vehicle. They perform two formulations of the system, first tenth-order vehicle model, and second simplified bicycle model. \cite{hu2020lane} has formulated an approach that combines the lane detection and model predictive control for accurate and stable control of the autonomous vehicle. The vehicle dynamics model consists of three degrees of freedom, and the fuzzy controller corrects the wheel steering angle .\cite{tosolin2020development} introduces model predictive control higher-level controller in combination with a low-level controller generating steering wheel angle and drive pedal position for the autonomous vehicle. \cite{alcala2018autonomous} introduces the Lyapunov-based approach with a Linear-Quadratic Regulator-Linear Matrix Inequality (LQR-LMI) tuning for control of the autonomous vehicle. The vehicle is modeled as a kinematic bicycle model. The nonlinear Lyapunov controller's parameters are optimized using a closed-loop system in linear parameter varying (LPV) form. 
\par
An autonomous vehicle (Sandstrom) uses a PID controller's variant as a feedback controller to regulate position and system actuators. They have used a simple geometric model for steering and speed control \cite{urmson2006robust}. Stanley, the autonomous vehicle which won the DARPA challenge, develops a control layer that regulates the steering, brake, and throttle. Two closed-loop Proportional-Integral (PI) trajectory tracking controllers are developed, one for steering control and other for throttle and brake \cite{thrun2006stanley}. Boss has used model-predictive control to perform vehicle control\cite{urmson2008autonomous}. Baidu Apollo uses a combination of three controllers, two low-level controllers consisting of feed-forward PID controller and LQR controller. A model- predictive controller is used as a higher-level controller to optimize PID and LQR parameters \cite{huang2018apolloscape}.

\subsection{Prediction of steering angle and speed from images}
In $1989$, Pomerleau has made the first attempt to drive the car, Autonomous Land Vehicle in a Neural Network (ALVINN), using the information from images and laser finder \cite{pomerleau1989autonomous}.  (ALVINN) has demonstrated the autonomous steering of a vehicle on public roads.  A simple neural network with $3$ layers is trained in an end-to-end fashion, which shows the potential of ANN for autonomous navigation. Inspired by ALVINN, recently, a team of NVIDIA engineers has trained an end-to-end convolution neural network (CNN) that processes images from a front-facing camera \cite{bojarski2016end}. The CNN learns the useful road features and predicts the steering commands. Du et al. explore two different frameworks to predict steering angle accurately and robustly using image data \cite{du2019self}. In the framework, the first network uses a $3$D convolution layer combined with the LSTM unit's recurrent layers. The second network incorporates transfer learning to achieve high-quality results. Lu et al. have proposed an end-to-end neural model trained with temporal and spatial visual cues \cite{chi2017deep}. The framework is composed of the feature extracting sub-network and steering-prediction sub-network. The feature sub-network includes spatio-temporal convolution, multi-scale residual aggregation, and convolutional LSTM, followed by ReLU and dropout layers for extracting the features. The steering-prediction network includes LSTM for predicting the steering angle. Similarly, Chowdhuri et al. have used the behavior modalities and images to develop a multi-task and a multi-model Z2Color network \cite{chowdhuri2019multinet}. Yang et al. have predicted the steering angle and speed control using a multi-modal framework, which took information from previous speed and visual data \cite{yang2018end}.

\par
 Kim et al. have proposed a two-stage network \cite{kim2017interpretable}. The first stage consists of an end-to-end network that predicts the steering angle from images, and the second stage generates attention heat map from images, which emphasizes on the regions in the images that influence the output of the network. Xu et al. have introduced a novel FCN-LSTM network that jointly learns from steering and segmentation loss \cite{xu2017end}. Hou et al. have proposed the FM-Net for the prediction of steering angle by incorporating the heterogeneous auxiliary networks for learning the steering angle \cite{hou2019learning}. In their work, they have used PSPNet \cite{zhao2017pyramid} and Flownet \cite{ilg2017flownet} as an auxiliary network that provides the feature maps at three different levels; low, medium, and high for feature mimicking to train the network in an end-to-end manner for steering angle prediction.

\section{Proposed Method}

\begin{figure*}[t]
      \centering
      \includegraphics[width=18cm]{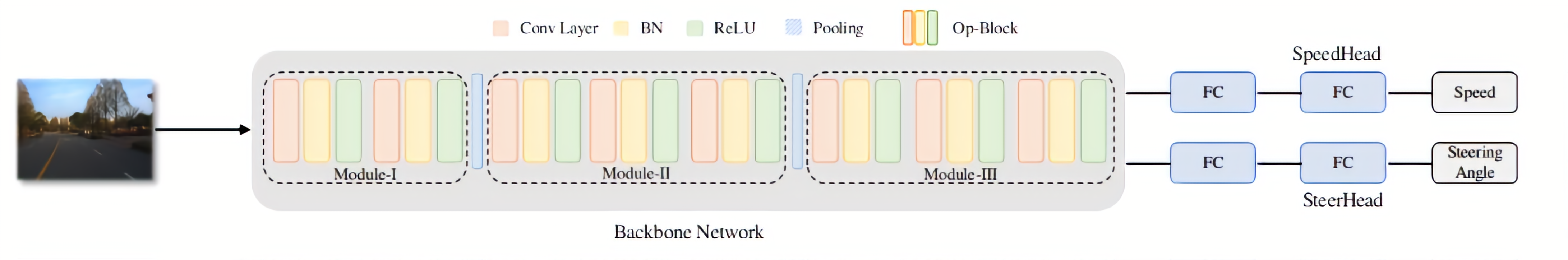}
      \caption{The architecture for predicting steering angle and speed using image data. The architecture consists of a backbone network and the separate heads for both steering angle and speed prediction. The backbone network consists of three modules, followed by SpeedHead and SteerHead. }
      \label{framework2}
\end{figure*}

\subsection{{$N^2$}C: Neural Network Controller Design}
The neural network has been widely used as a function approximation for classification and regression problems. Besides the feed-forward neural network, recurrent neural network (RNNs) solves the sequence modeling problems by having recurrent connections between nodes of the network. RNNs with long short-term memory (LSTM) \cite{gers2002learning} units tend to capture sequential data dependencies, providing leverage to model the sequential structure of CAN bus data. 
\par 

\begin{figure}[b]
      \centering
      \includegraphics[width=9cm]{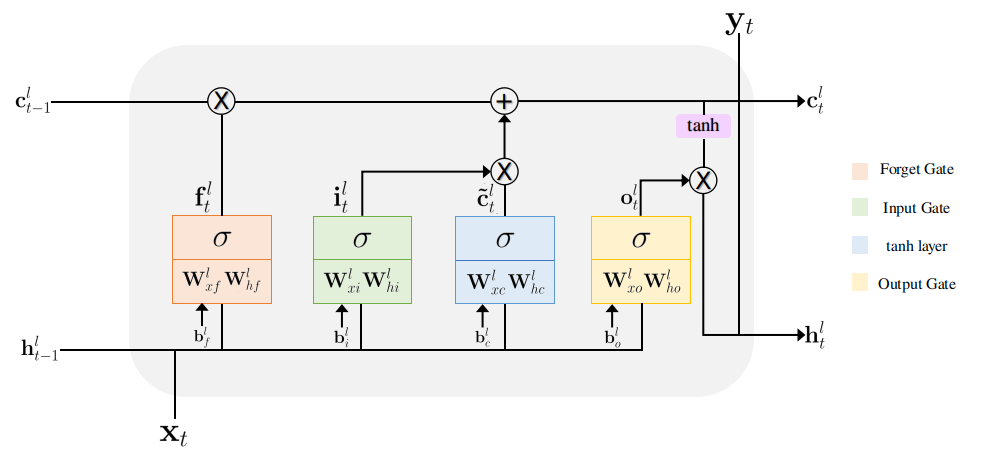}
      \caption{The schematic of lstm cell which includes the input gate, output gate, forget gate and cell state memory.}
      \label{lstm-cell}
\end{figure}

The standard deep LSTM architecture is the combination of input layers, LSTM layers, fully connected layers, and output layers. In the LSTM layers, the basic building block is LSTM cells that map the input sequence layer to the output sequence layer. Fig. \ref{lstm-cell} shows the basic LSTM cell, which comprises of four interacting units: an internal cell, a forget gate, an input gate, and an output gate. The internal cell state at the previous time step is memorized at the internal cell through recurrent connections. The flow of input activation to the internal cell is controlled by the input cell. The forget gate is responsible for forgetting and resetting the cell memory adaptively, and an output layer controls the output activation flow to the LSTM cell. Suppose the input state to the LSTM cell is ${x_t}$, the input gate, forget gate, output gate, hidden state output and cell state memory are denoted by ${i^l_t}$,${f^l_t}$, ${o^l_t}$, ${h^l_t}$ and ${c^l_t}$ respectively at the time step $t$ for $l^{th}$ LSTM network layer. In addition, at time step $t-1$ the hidden state output and cell memory state are represented by ${h^l_{t-1}}$ and ${c^l_{t-1}}$ respectively. The governing equation from Eq. (\ref{1})-Eq. (\ref{6}) shows the relationship between the variables, as mentioned earlier.

\begin{flalign}
\label{1}
\hspace{0.7cm}
\bold{i}_{t}^{l} = \sigma (\bold{W}_{xi}^{l}\bold{x}_t+\bold{W}_{hi}^{l}\bold{h}_{t-1}+\bold{b}_i^l) \; , &&
\end{flalign}

\begin{flalign}
\label{2}
\hspace{0.7cm}
\bold{f}_{t}^{l} = \sigma (\bold{W}_{xf}^{l}\bold{x}_t+\bold{W}_{hf}^{l}\bold{h}_{t-1}+\bold{b}_f^l)\; , &&
\end{flalign}

\begin{flalign}
\label{3}
\hspace{0.7cm}
\bold{o}_{t}^{l} = \sigma (\bold{W}_{xo}^{l}\bold{x}_t+\bold{W}_{ho}^{l}\bold{h}_{t-1}+\bold{b}_o^l)\; , &&
\end{flalign}

\begin{flalign}
\label{4}
\hspace{0.7cm}
\bold{\tilde{c}}_{t}^{l} = \tanh (\bold{W}_{xc}^{l}\bold{x}_t+\bold{W}_{hc}^{l}\bold{h}_{t-1}+\bold{b}_c^l)\; , &&
\end{flalign}

\begin{flalign}
\label{5}
\hspace{0.7cm}
\bold{c}_{t}^{l} = \bold{f}_{t}^{l} \odot \bold{c}_{t-1}^{l} + \bold{i}_{t}^{l} \odot \bold{\tilde{c}}_{t}^{l}\; , &&
\end{flalign}

\begin{flalign}
\label{6}
\hspace{0.7cm}
\bold{h}_{t}^{l} = \bold{o}_{t}^{l} \odot \tanh(\bold{c}_{t}^{l})\; , &&
\end{flalign}
where $\bold{W}$ and $\bold{b}$ represent the weight matrix and bias vector corresponding to different inputs within different gates.  
\par 
In this study, a sequence to sequence LSTM network is used having multiple LSTM layers for the design of neural network based controller as illustrated in Fig. \ref{framework}. It includes the input layer, five LSTM layers, one fully connected layer (FC), and the output layer. The number of units in the five LSTM layers are denoted by $l_1$, $l_2$, $l_3$, $l_4$ , and $l_5$ respectively. $v$ represents the number of nodes in the FC layer. The input and output data for training the architecture is formatted in three dimensional where $r$ shows the sample entries in the first dimension and $n$ as a second dimension corresponds to time steps. The stack of five LSTM layers are used in the proposed network and based on that it is referred as $LSTM-5$.  Algorithm-1 illustrates the pseudo-code for the $N^2C$. The input to $LSTM-5$ are linear velocity and steering angle. The input data is collected through behavioral cloning by driving the vehicle manually. The behavioral cloning is defined as learning from observations, in which the human observations are recorded along with actions. This human observation and action pairs are used by the learning agent to simulate the human behavior for that particular task. In this work, the human observations in term of throttle( drivepedal), brake (brakepedal) and torque along with linear velocity, and steering angle are recorded. The importance of using the behavioral cloning in this work, is to learn the human driving by observation and optimize the control of the autonomous vehicle through a learning algorithm in this case $LSTM-5$ for smoothness of autonomous driving. The output of $LSTM-5$ is a prediction of throttle, brake and torque. Let us denote the input to the $LSTM-5$ as $\bold{X} = {\bold{x_1},\bold{x_2},\bold{x_3},...,\bold{x_n}}^T \in \mathbb{R}^{n \times u}$ and output as  $\bold{Y} = {\bold{y_1},\bold{y_2},\bold{y_3},...,\bold{y_n}}^T \in \mathbb{R}^{n \times v}$  with $n$ time steps. The input and output features are represented by $u$ and $v$ respectively. The rows in the both $\bold{X}$ and $\bold{Y}$ corresponds to time steps and the columns for features. The $LSTM-5$ takes the input sequence $\bold{X}$ and maps it for the prediction of output sequence $\bold{Y}$ over the full duration of time steps $(t = 1,2,3,...,n)$ with each LSTM cell in the LSTM layer is connected to its two neighbours $(t-1,t,t+1)$. The network is trained by continually sending the $\bold{x_t}$ to the $LSTM-5$ over the entire temporal space using the repeated LSTM cells. At the end of the network, the fully connected layer (FC) is used to interlink the LSTM and target output layers in order to construct the desired output.   

\begin{algorithm}[h]
\SetAlgoLined
\KwIn{Linear Velocity and Steering Angle: $\bold{X} = {\bold{x_1}, \bold{x_2}, \bold{x_3} ,..., \bold{x_n}}^T \in \mathbb{R}^{n \times u}$}
\KwResult{Prediction of Throttle, Brake and Torque: ${Y}$}
\bf Given parameters: $\bold{W}_{xi}$, $\bold{W}_{hi}$, $\bold{W}_{xf}$, $\bold{W}_{hf}$, $\bold{W}_{xo}$, $\bold{W}_{ho}$, $\bold{W}_{xc}$, $\bold{W}_{hc}$, $\bold{b}_{i}$, $\bold{b}_{f}$, $\bold{b}_{o}$, $\bold{b}_{c}$ \\
Initialization of $\bold{h}_{0}$, $\bold{c}_{0}$ = $\vec{0}$\\ 
\While {t $	\leq$ M}{
Calculate $\bold{{i}_{t}}$ using Eq. \ref{1}\\
Calculate $\bold{{f}_{t}}$ using Eq. \ref{2}\\
Calculate $\bold{\tilde{c}}_{t}$ using Eq. \ref{4}\\
Update the cell state ($\bold{c}_{t}$) using Eq. \ref{5}\\
Calculate $\bold{o}_{t}$ using Eq. \ref{3}\\
Calculate $\bold{h}_{t}$ using Eq. \ref{6}\\
t $\leftarrow$ t$+1$
}
\KwOut{h = $[\bold{h_1}, \bold{h_1}, ..., \bold{h_n}]$}
Y = $tanh(h)$  Fully Connected layer  \\
\KwResult{Prediction of Throttle, Brake and Torque: $\bold{Y} = {\bold{y_1}, \bold{y_2}, \bold{y_3}, ..., \bold{y_n}}^T \in \mathbb{R}^{n x v}$}
% \procedure LSTM-5 (
% \textit{$P_i$}:Input point cloud data\\
% \textit{$P_s$}:Point set of point cloud\\
% \hspace{2mm}$P_s$:{$p_1,p_2,p_3,...,p_k$}  $k=1,2,3,...,n$ \\
% \textit{$P_r$}:Remaining points\\
% \textit{$A_t$}:Angle threshold\\
% \textit{$D_t$}:Dynamic distance threshold\\
% \textit{$S_t$}:Size threshold\\
% \textit{$\theta$}:Angle between $p_k$ and $p_{k+1}$\\
% \hspace{2mm} where {$p_{k+1}$} is adjacent point\\
% \textit{$\delta$}:Distance between $p_k$ and $p_{k+1}$\\
% \hspace{2mm} where {$p_{k+1}$} is adjacent point\\
% \textit{$C$}:Candidate group\\
% \ForEach{$p_k$ $\in$ \textit{$P_s$}}{
% \eIf{$\theta$ < $A_t$}{
% Add ($p_k$) to the $C$
% }{
% \eIf{size of $C$ > $S_t$}{
% mark all the $p_k$ in $C$ as $P_g$
% }{
% mark all the $p_k$ in $C$ as $P_r$
% }
% } 
% }
% \ForEach{$p_k$ $\in$ \textit{$P_r$}}{
% \eIf{$\delta$ < $D_t$}{
% Add ($p_k$) to the $C$
% }{
% \eIf{size of $C$ > $S_t$}{
% mark all the $p_k$ in $C$ as vertical points
% }{
% mark all the $p_k$ in $C$ as $P_g$
% }
% } 
% }

\caption{$N^2C$ pseudo-code comprises of stacked LSTM layers. }
\end{algorithm}

% \begin{algorithm}[h]
% \KwIn{Linear Velocity and Steering Angle: $\bold{X} = {\bold{x_1},\bold{x_2},\bold{x_3},...,\bold{x_n}}^T \in \mathbb{R}^{n \times u}$}
% \KwOut{Throttle, Brake, and Torque: $\bold{Y} = {\bold{y_1},\bold{y_2},\bold{y_3},...,\bold{y_n}}^T \in \mathbb{R}^{n \times v}$ } 
% \nl Let M be dataset that include {$\bold{X},\bold{Y}$} \\
% \quad Apply data augmentation\\
% \nl \textbf{Initialization}:\\
% \quad \bf $l_i$: Number of units in LSTM layers, where $i=1,2,3,4,5$. \\
% \quad \bf $L$: Total number of layers in LSTM. \\
% \quad \bf $W$: Glorot uniform initialization for weights \\
% \quad \bf $d_r$: decay rate \\ 
% \quad \bf $l_r$: learning rate \\
% \nl \bf procedure compute-LSTM(M,$l_i$,W,$d_r$,$l_r$){

% }
% \nl \bf procedure LSTM-5(M,$l_i$,W,$d_r$,$l_r$){ \\
% \quad for $l$ $\leq$ $L$ \\
% \quad \quad compute-LSTM(M,W,$d_r$,$l_r$) \\

% }

% \nl \bf Pass\;
%     \caption{{\bf Algorithm} \label{Algorithm}}

% \end{algorithm}

\subsection{End-to-End Prediction of Steering Angle and Speed}

The second part of this study explores prediction of steering angle and speed using the images data with an end-to-end ANN, as shown in Fig. \ref{framework2}. A CNN is designed that takes image as input and predicts the steering angle and speed. The network is segregated into three parts i) backbone ii) SteerHead, and iii) SpeedHead. The backbone of the network extracts the features by passing the input image through a set of modules. In each module, a set of operations named op-block is performed on the input, which comprises convolution, batch normalization, and activation. A pooling layer is added after each module. In a particular module, the filter ($K$) size is increased by a factor of $2$ whereas, the same kernel size $F$ $(3$x$3)$ is used. The encoded features from the backbone architecture are parsed to two fully connected heads, SpeedHead, and SteerHead, for the prediction of speed and steering angle, respectively.  

% The second part of this study focuses on the prediction of steering angle and speed using the image data with an end-to-end ANN, as shown in Fig. \ref{framework2}. A CNN is designed that takes the images as input and predicts the steering angle and speed. The input image is down-sampled and passed through a set of operations, including convolution, batch normalization, and activation. In the proposed method, these sets of operations (convolution, batch normalization, and activation) are denoted as an operation block (op-block). The network architecture is divided into three parts i) backbone ii) SteerHead iii) SpeedHead. The network's backbone consists of three modules; each module consists of the op-block and is performed ahead by the pooling.   
\par 
% The input images are passed through the backbone, then into two fully connected heads, one for the prediction of speed and one for the steering angle prediction. In the proposed architecture, the SteerHead and SpeedHead represent the prediction of steering angle and speed, respectively. In each of these heads, batch normalization is conducted, and a rectified non-linear unit (ReLU) is applied. 

% \usepackage{booktabs}
% \usepackage{graphicx}
\begin{table*}[]
\centering
\caption{RMSE and MAE scores of the proposed method using mse as loss function on the collected dataset. }
\label{table-1}
\resizebox{12cm}{!}{%
\begin{tabular}{@{}c|cc|cc|cc|cc|cc@{}}
\toprule
\multicolumn{1}{l|}{} &
  \multicolumn{2}{c|}{Human Driving} &
  \multicolumn{2}{c|}{PP+NN} &
  \multicolumn{2}{c|}{PP+PID} &
  \multicolumn{2}{c|}{MPC+NN} &
  \multicolumn{2}{c}{MPC+PID} \\ \midrule
Networks       & RMSE    & MAE    & RMSE            & MAE             & RMSE   & MAE    & RMSE            & MAE             & RMSE   & \multicolumn{1}{c}{MAE} \\ \midrule
LSTM-5 Network & 0.05593 & 0.032  & 0.0596 & 0.0429 & 0.1075 & 0.0664 & 0.057  & 0.0367 & 0.0802 & 0.0492                   \\
Conv-LSTM      & 0.0643  & 0.0381 & 0.0796 & 0.0477 & 0.1075 & 0.0664 & 0.065  & 0.0425 & 0.0802 & 0.0492                   \\
Conv           & 0.0683  & 0.0421 & 0.0958 & 0.0511 & 0.1075 & 0.0664 & 0.0697 & 0.0493 & 0.0802 & 0.0492                   \\ \bottomrule
\end{tabular}%
}
\end{table*}

% \usepackage{booktabs}
% \usepackage{graphicx}
\begin{table*}[]
\centering
\caption{RMSE and MAE scores of the proposed method using smooth-L1 as loss function on the collected dataset.}
\label{table-2}
\resizebox{12cm}{!}{%
\begin{tabular}{@{}c|ll|ll|ll|ll|ll@{}}
\toprule
\multicolumn{1}{l|}{} &
  \multicolumn{2}{c|}{Human Driving} &
  \multicolumn{2}{c|}{PP+NN} &
  \multicolumn{2}{c|}{PP+PID} &
  \multicolumn{2}{c|}{MPC+NN} &
  \multicolumn{2}{c}{MPC+PID} \\ \midrule
Networks &
  \multicolumn{1}{c}{RMSE} &
  \multicolumn{1}{c|}{MAE} &
  \multicolumn{1}{c}{RMSE} &
  \multicolumn{1}{c|}{MAE} &
  \multicolumn{1}{c}{RMSE} &
  \multicolumn{1}{c|}{MAE} &
  \multicolumn{1}{c}{RMSE} &
  \multicolumn{1}{c|}{MAE} &
  \multicolumn{1}{c}{RMSE} &
  \multicolumn{1}{c}{MAE} \\ \midrule
LSTM-5 Network &
  \textbf{0.04963} &
  \textbf{0.03403} &
  \textbf{0.0547} &
  \textbf{0.0419} &
  0.1075 &
  0.0664 &
  \textbf{0.0515} &
  \textbf{0.0355} &
  0.0802 &
  0.0492 \\
Conv-LSTM &
  0.0558 &
  0.037 &
  0.0789 &
  0.0461 &
  0.1075 &
  0.0664 &
  0.0598 &
  0.0373 &
  0.0802 &
  0.0492 \\
Conv &
  0.0624 &
  0.04 &
  0.0854 &
  0.0491 &
  0.1075 &
  0.0664 &
  0.0653 &
  0.0402 &
  0.0802 &
  0.0492 \\ \bottomrule
\end{tabular}%
}
\end{table*}

\section{Experimentation and Results}

\subsection{CAN Data Extractor}
Real-time data is communicated between different electronic components in a vehicle through a standard CAN bus. In this study, a CAN bus data extractor is designed for the KIA Soul EV (our research testbed) that contains i) can-shield and ii) drivekit. The can-shield uses MCP2515 CAN bus controller with the SPI interface along with MCP2551 CAN transceiver. The can-shield provides the odometry data through the OBD-II CAN network at the speed of $1$Mb/sec. 
\par 
The drivekit\footnote{https://github.com/PolySync/oscc} serves as an electronic controller for the drive-by-wire vehicles that include the KIA Soul EV. The drivekit allows us to send and receive control commands through the OBD-II CAN network, which includes the vehicle's state information. The drivekit has its own CAN bus called Control CAN, which communicates with OBD-II CAN network through CAN Gateway. The communication through this mechanism ensures the safety of the vehicle's CAN bus. 
\par 
Besides the hardware integration, the CAN messages are decoded using a can-sniffer\footnote{https://github.com/linux-can/can-utils} algorithm in which the CAN arbitration identifiers are determined heuristically, by filtering out the specific arbitration identifiers corresponds to speed, steering angle, throttle, brake, and torque. The ROS node for can-shield and drivekit is used for extracting the data from both CAN data extractor mechanism. 

\subsection{Datasets}
% In this work, we have collected our own dataset which includes the CAN bus and image data. In addition, we have also used Udacity dataset for the evaluation of the proposed method. The CAN bus data in our dataset includes throttle, brake, torque, velocity and steering angle which are obtained through the vehicle CAN bus using behavior cloning while driving vehicle manually. The throttle, brake values are normalized between $[0,1]$ whereas, the torque values are between $[-1,1]$. The velocity values are recorded in $m/s$ and steering angle values are in radians. The CAN data is collected for two scenerios where one scene CAN data is used for training and other one is used for the testing the proposed method. The train CAN data consist of $50,000$ sequences whereas the test CAN data consist of $17000$ sequeces. Both the CAN data are recorded using our testbed as shown in Fig-1. In recording the CAN bus data there is no restriction of speed limit and the driver is free to drive the car in any manner, only by abiding the traffic rules. The image data is collected using our testbed that is composed of video frames. The image data and CAN bus data is synchronized and annotated with the respective steering angle and velocity. In collecting the image data only single front camera is used having the resolution of $1440 \times 1080$.
We used our research testbed with perceptional and navigational sensors for the data collection, as shown in Fig. \ref{car}. The perception sensor suite includes a $32$ and $16$ channels Velodyne lidar along with two FLIR BlackFly S cameras having a USB interface. Novatel GNSS is used as a navigational sensor for localization and mapping of the autonomous vehicle. The CAN data is extracted through drivekit and can-shield.
\par 
In our experimental study, only CAN bus and image data are collected. The CAN bus data includes throttle, brake, torque, speed, and steering angle, obtained through the vehicle CAN bus using behavioral cloning while driving the vehicle manually. The throttle, brake values are normalized between $[0,1]$ whereas, the torque values are between $[-1,1]$. The speed values are recorded in $m/s$, and steering angle values are in radians. The CAN data is collected for two scenarios where one scene CAN data is used for training, and the other is used to test the proposed method. The training CAN data consists of $50,000$ sequences, whereas the test CAN data consists of $17000$ sequences. In recording the CAN bus data, there is no restriction of the speed limit, and the driver is free to drive the car in any manner, only by abiding the traffic rules. The image data is collected using our testbed that is composed of video frames. The image data and CAN bus data is synchronized and annotated with the respective steering angle and speed. In collecting the images data, the single front camera is used, having a resolution of $1440 \times 1080$. 
\par 
We have also used the Udacity\footnote{https://github.com/udacity/self-driving-car} dataset for the evaluation of the proposed method. The reason for using the Udacity dataset in contrast to other datasets, for instance, Comma.ai\footnote{https://github.com/commaai/comma2k19}, is that it includes CAN bus and image data which is advisable for our experimentation. In Comma.ai for instance, only image data with the corresponding steering angle is provided, and there is no speed information is available, irrespective of this, the Udacity dataset comprises of latitude, longitude, gear, brake, throttle, steering angles, speed, and image data.

% In comparison to other datasets like Comma.ai, which only provides images along with steering angle and speed is not applicable to use for our work. The Udacity dataset includes 

\subsection{{$N^2$}C Controller Design}

This section focuses on implementing proposed $LSTM-5$ architecture for the throttle, brake, and torque prediction, using human driving data of speed and steering angle. For the useful comparison of our proposed method, both the collected and Udacity datasets are used.
\par 
The training data consist of speed, steering angle, throttle, brake, and torque. The input to the network includes speed and steering angle whereas, throttle, brake, and torque serve as the output of the network. In the proposed method, both front and rear wheels speed are collected from the vehicle CAN bus. In training the neural network, all the wheel speed are averaged and used for the experimental evaluation. In order to train the $LSTM-5$ network, both the input and output are converted into three-dimensional arrays. The first dimension entries correspond to samples, the second dimension represents the timestep, and the third dimension shows the input/output features/channels. Both input and output are fed to the $LSTM-5$ for the training of the model after converting it to an input format. In addition, the training data is augmented with flipping, warping, and noise injection to enhance the trained network's generalization without being over-fitted to the train data. In training, the glorot uniform initializer is used for randomly initializing the weights of all LSTM layers and the FC layer, respectively. Besides, Adam (Adaptive Momentum Estimation) is used as an optimizer with a learning rate of $0.008$, having a decay rate of $0.0001$.  The number of LSTM cells in the $LSTM-5$ network layers are $(300,250,100,50,20)$ and this configuration is chosen empirically. The training process runs for a total of $500$ epochs, with the batch size of $32$ using the Keras and Tensorflow deep learning library on the Nvidia GTX 1080Ti GPU card. In our case, the two different loss function is used for the experimentation i) mean squared error (MSE) ii) Huber loss (smooth-L1 loss). The mean squared error (MSE) and Huber loss  (smooth-L1 loss) are mathematically given by Eq. (\ref{7}) and Eq. (\ref{8}).

\begin{flalign}
\label{7}
\hspace{0.7cm}
MSE = \frac{1}{k}\sum_{i=1}^{k}(y_k-\hat{y_k})^2 \; , &&
\end{flalign}

\begin{flalign}
\label{8}
\hspace{0.7cm}
L_{\delta}(x) = 
\left\{\begin{matrix}
\frac{1}{2}x^{2} \;&&for \left | x \right | \leq \delta \\
\delta(\left | x \right | - \frac{1}{2}\delta ), \;&& otherwise 
\end{matrix}\right. \;  &&
\end{flalign}
where in Huber loss, the value of $\delta=1.0$ is used in experimentation.

\subsubsection{Analysis}
The proposed method's effectiveness is determined by evaluating the trained model $N^2C$  on the collected test data and Udacity data. It is also analyzed with a PID controller in combination with the path tracking algorithm.  The path tracking algorithms used in the analysis are pure pursuit (PP) and model predictive control (MPC) based path follower. These path tracking algorithms are implemented and tested on testbed along with the PID controller. The angular velocity from the path tracking algorithm is converted to the steering angle before evaluation. For evaluation metric, root mean squared error (RMSE) and mean absolute error (MAE) are used. Table. \ref{table-1} , and Table. \ref{table-2} shows the evaluation results of $N^2C$ with PID along with pure pursuit and mpc based path follower on the collected test dataset. The smaller RMSE and MAE values correspond to better accuracy. The RMSE and MAE scores are evaluated using both loss functions used in training the proposed model. RMSE and MAE are evaluated using the Eq. (\ref{9}) and Eq. (\ref{10}), respectively. 
 \par
 In Table. \ref{table-1} , and Table. \ref{table-2}, the human driving (proposed method only evaluated on human driving data) RMSE and MAE scores are evaluated only with the collected test data without any path tracking algorithm. Also, the proposed method, when used with path tracking algorithms (i-e PP+NN and MPC+NN) have better RMSE and MAE scores as compared to the PID controller with path tracking algorithms (PP+PID and MPC+PID). Fig. \ref{analysis-1} shows the individual throttle, brake, and torque RMSE and MAE scores of the proposed method for both loss functions only for human driving using collected test data. The individual comparison of the throttle, brake, and torque from the proposed method and PID along with pure pursuit and mpc based path follower is represented in Fig. \ref{analysis-2}. 
 \par 
 The proposed method is also evaluated using the Udacity dataset. Table-\ref{table-3} shows the RMSE and MAE scores for the Udacity dataset. In this analysis, we have not experimented with the effect of the path tracking algorithm because, in the Udacity dataset, there is no information about the linear and angular velocities from the path tracking algorithm.
 \par 
 The $N^2C$ is designed by training the neural network on data acquired from the vehicle CAN bus. The environmental conditions does not play any role in the prediction of controller output. In addition, the traffic condition are incorporated in the perception and planning modules of our autonomous vehicle stack. In this work, we have explicitly presented the neural network-based controller for the autonomous vehicle using behavioral cloning to provide smoothness to the control of autonomous vehicle. In addition, the proposed $N^2C$ can be adapted for the other vehicles by replacing the CAN data extractor with the required vehicle CAN data extractor module. The fine tuning based on the $N^2C$ will also be deployed for any other vehicle.
 
 \subsubsection{Effect of Noise on $N^2C$}
 In this section, the neural network-based controller's performance for the effect of noise in case of inaccurate input data is evaluated. For the evaluation of controller, a Gaussian noise models having mean($\mu=0$), and sigma($\sigma=1$) is utilized. The experimentation is performed on test data for both the collected and Udacity datasets. Table-\ref{noise} illustrates the RMSE and MAE scores for the effect of noise on the controller output.
 
% \usepackage{booktabs}
% \usepackage{graphicx}
\begin{table}[]
\centering
\caption{Quantitative evaluation of effect of noise on $N^2C$ for collected and Udacity datasets.}
\label{noise}
\resizebox{9cm}{!}{%
\begin{tabular}{@{}ccccccccc@{}}
\toprule
\multicolumn{9}{c}{Human Driving}                                             \\ \midrule
\multicolumn{5}{c|}{Collected Dataset} & \multicolumn{4}{c}{Udacity Dataset} \\ \midrule
\multicolumn{1}{c|}{Loss Function} &
  \multicolumn{2}{c|}{Smooth-L1} &
  \multicolumn{2}{c|}{MSE} &
  \multicolumn{2}{c|}{Smooth-L1} &
  \multicolumn{2}{c}{MSE} \\ \midrule
\multicolumn{1}{c|}{Networks} &
  RMSE &
  \multicolumn{1}{c|}{MAE} &
  RMSE &
  \multicolumn{1}{c|}{MAE} &
  RMSE &
  \multicolumn{1}{c|}{MAE} &
  RMSE &
  MAE \\ \midrule
\multicolumn{1}{c|}{LSTM-5 Network} &
  0.0505 &
  \multicolumn{1}{c|}{0.0351} &
  0.0612 &
  \multicolumn{1}{c|}{0.0332} &
  0.0578 &
  \multicolumn{1}{c|}{0.0382} &
  0.0654 &
  0.0367 \\
\multicolumn{1}{c|}{Conv-LSTM} &
  0.0617 &
  \multicolumn{1}{c|}{0.0382} &
  0.0664 &
  \multicolumn{1}{c|}{0.0391} &
  0.0632 &
  \multicolumn{1}{c|}{0.0396} &
  0.0778 &
  0.04012 \\
\multicolumn{1}{c|}{Conv} &
  0.0637 &
  \multicolumn{1}{c|}{0.0422} &
  0.071 &
  \multicolumn{1}{c|}{0.0453} &
  0.0721 &
  \multicolumn{1}{c|}{0.0479} &
  0.081 &
  0.0496 \\ \bottomrule
\end{tabular}%
}
\end{table}

 \begin{flalign}
\label{9}
\hspace{0.7cm}
RMSE = \sqrt{\frac{1}{k}\sum_{j=1}^{k}(y_k-\hat{y_k})^2} \; , &&
\end{flalign}

 \begin{flalign}
\label{10}
\hspace{0.7cm}
MAE = \frac{1}{k}\sum_{j=1}^{k}\left \| y_k-\hat{y_k} \right \| \; , &&
\end{flalign}

\begin{figure*}[t]
      \centering
      \includegraphics[width=13cm,height=4cm]{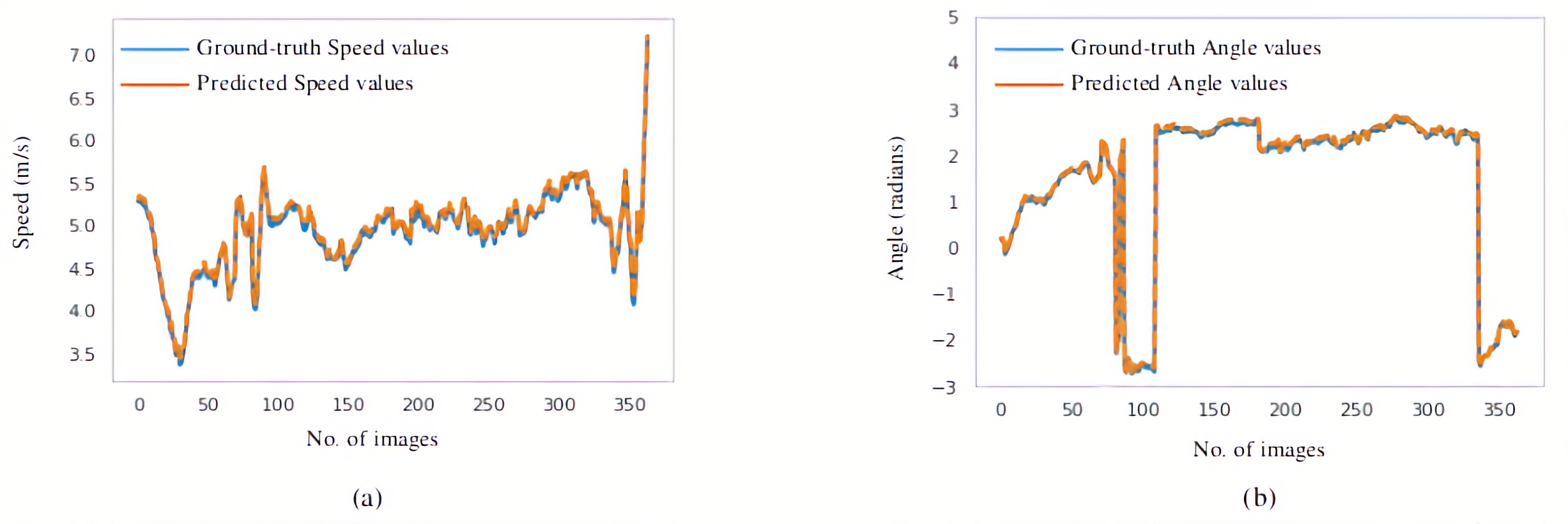}
      \caption{Quantitative comparison between the ground-truth and predictive values for (a) speed, and (b) angle on the collected dataset. }
      \label{qualitative-ours}
\end{figure*}

\begin{figure*}[t]
      \centering
      \includegraphics[width=13cm,height=4cm]{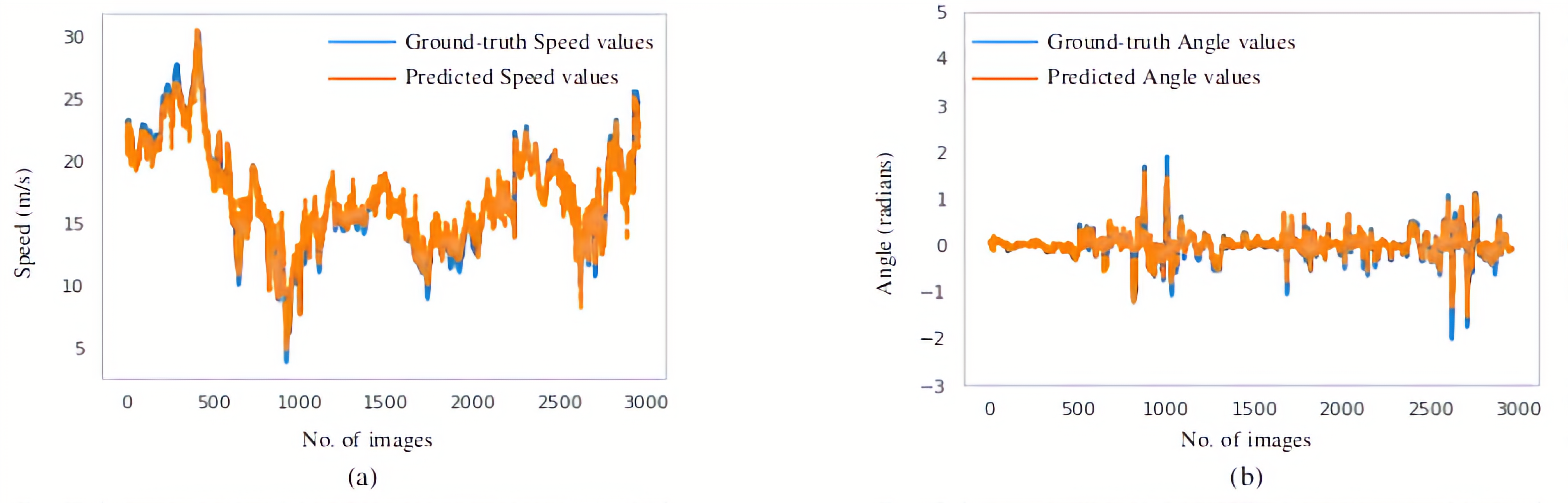}
      \caption{Quantitative comparison between the ground-truth and predictive values for (a) speed, and (b) angle on Udacity dataset. }
      \label{qualitative-udacity}
\end{figure*}

% \begin{figure*}[h]
% \begin{subfigure}{0.5\textwidth}
% \centering
% \includegraphics[width=0.9\linewidth, height=3.5cm]{images/velocity_profile_ours.png} 
% \caption{}
% \label{velocity-profile-ours}
% \end{subfigure}
% \quad
% \begin{subfigure}{0.5\textwidth}
% \centering
% \includegraphics[width=0.9\linewidth, height=3.5cm]{images/angle_profile_ours.png}
% \caption{}
% \label{angle-profile-ours}
% \end{subfigure}
% \quad

% \caption{Quantitive comparison between the ground-truth and predictive values for (a) velocity and (b) angle on the collected dataset. }
% \label{qualitative-ours}
% \end{figure*}

% \begin{figure*}[h]
% \begin{subfigure}{0.5\textwidth}
% \includegraphics[width=0.9\linewidth, height=3.5cm]{images/velocity_profile_udacity.png} 
% \caption{}
% \label{velo-udacity}
% \end{subfigure}
% \quad
% \begin{subfigure}{0.5\textwidth}
% \includegraphics[width=0.9\linewidth, height=3.5cm]{images/angle_profile_udacity.png} 
% \caption{}
% \label{angle-udacity}
% \end{subfigure}

% \caption{Quantitive comparison between the ground-truth and predictive values for (a) velocity and (b) angle on Udacity dataset.}
% \label{qualitative-udacity}
% \end{figure*}

% \usepackage{graphicx}
% \usepackage{booktabs}
% \usepackage{graphicx}
\begin{table}[h]
\centering
\caption{RMSE and MAE scores of LSTM-5, Conv-LSTM and Conv network with smooth-L1  and mse loss function on Udacity dataset.}
\label{table-3}
\resizebox{7cm}{!}{%
\begin{tabular}{@{}ccccc@{}}
\toprule
\multicolumn{5}{c}{Human Driving}                                                                    \\ \midrule
\multicolumn{1}{c|}{Loss Function} & \multicolumn{2}{c|}{Smooth-L1}        & \multicolumn{2}{c}{MSE} \\ \midrule
\multicolumn{1}{c|}{Networks}      & RMSE    & \multicolumn{1}{c|}{MAE}    & RMSE        & MAE       \\ \midrule
\multicolumn{1}{c|}{LSTM-5 Network} & \textbf{0.05314} & \multicolumn{1}{c|}{\textbf{0.0315}} & \textbf{0.06253} & \textbf{0.0356} \\
\multicolumn{1}{c|}{Conv-LSTM}     & 0.05834 & \multicolumn{1}{c|}{0.0325} & 0.07153     & 0.0421    \\
\multicolumn{1}{c|}{Conv}          & 0.06647 & \multicolumn{1}{c|}{0.0424} & 0.07541     & 0.0489    \\ \bottomrule
\end{tabular}%
}
\end{table}
\subsubsection{Ablation Study}
% In order to make a constructive analysis, addition to proposed model $N^2C$, we have analyzed different adjacent configuration in designing the neural network controller based on human driving. First we tried with Single-Layer LSTM having $300$ LSTM cell in that single layer. The training configuration is kept similar to the proposed method. This Single-Layer LSTM model is evaluation on test data as well as compared with PID along with path tracking algorithm that includes pure pursuit and mpc based path follower. Table- shows the quantitative results of Single-Layer LSTM network.
The proposed model, $N^2C$, is analyzed with different adjacent configurations in designing the neural network controller based on behavioral cloning. The experimentation is performed with a convolution neural network. Two networks are designed, including a single convolution layer and a Conv-LSTM network with a $1D$ convolution layer. In the first network, the five $1D$ convolution layers are used with ReLU as an activation function. After every two layers, dropout layers are used to regularize the training to reduce the over-fitting. The last dense layer of the network uses a $tanh$  activation function so that the output values are sampled in the range of $[-1,1]$. For the Conv-LSTM network, $1D$ convolution is applied to input data and then passed to the LSTM network. The network configuration of this network includes two $1D$ convolution layers, followed by four LSTM layers. The dropout layers are used after convolution layers, to avoid the overfitting.  The mse and smooth-L1 Loss functions are used for training both of the networks. 
\par
\begin{figure*}[b]

\begin{subfigure}{0.5\textwidth}
\includegraphics[width=7cm]{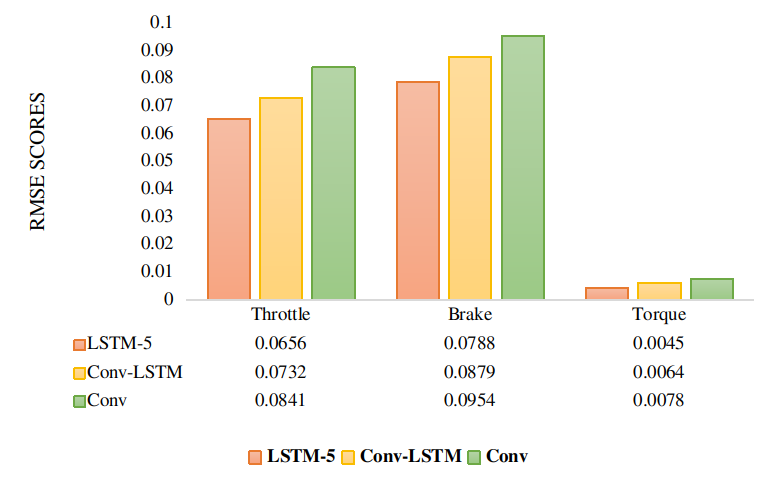} 
\caption{}
\label{HD-smooth-L1-RMSE}
\end{subfigure}
\quad
\begin{subfigure}{0.5\textwidth}
\includegraphics[width=7cm]{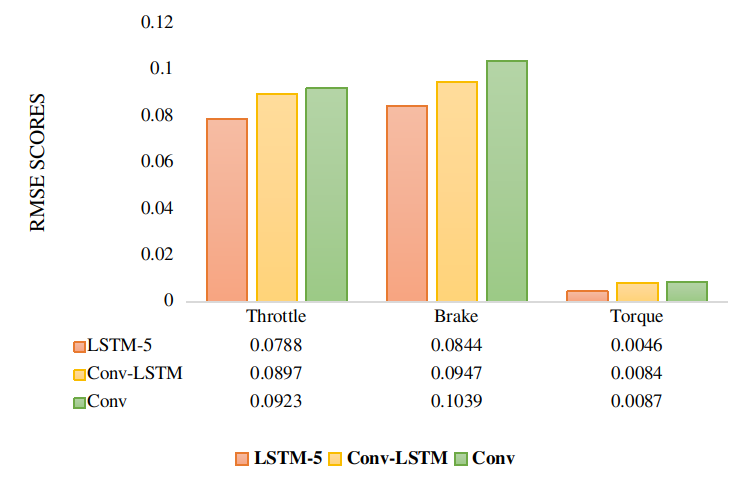}
\caption{}
\label{HD-mse-RMSE}
\end{subfigure}
\quad
\begin{subfigure}{0.5\textwidth}
\includegraphics[width=7cm]{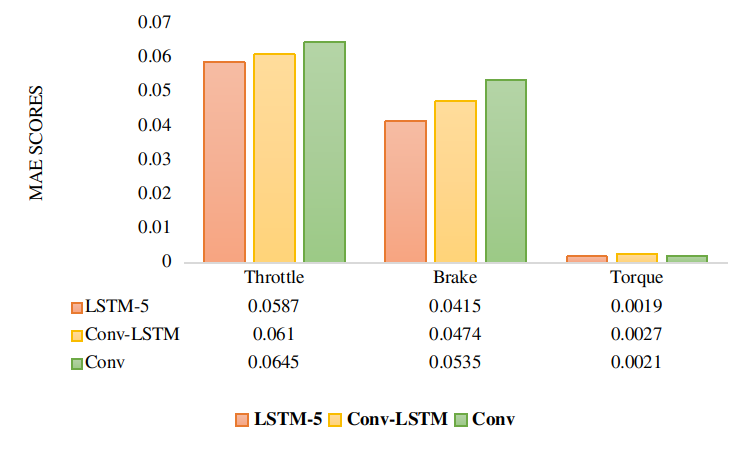} 
\caption{}
\label{HD-smooth-L1-MAE}
\end{subfigure}
\quad
\begin{subfigure}{0.5\textwidth}
\includegraphics[width=7cm]{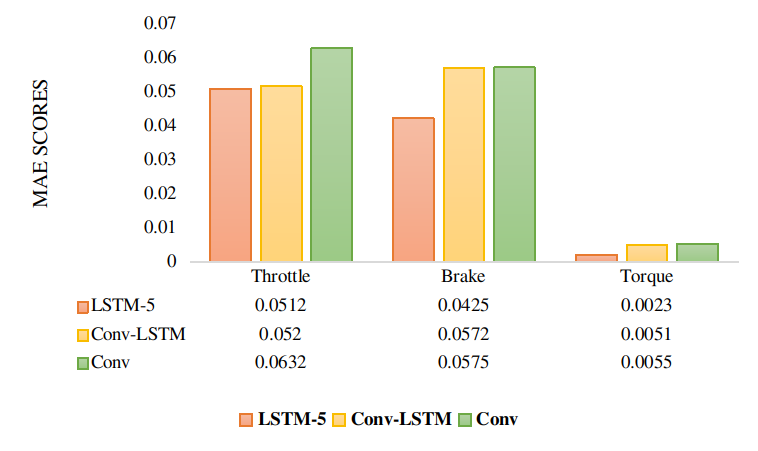} 
\caption{}
\label{HD-mse-L1-MAE}
\end{subfigure}

\caption{Individual throttle, brake, and torque RMSE and MAE scores on human driving. (a),(c) RMSE and MAE scores of the throttle, brake, and torque for LSTM-5, Conv-LSTM, and Conv network using smooth-L1 loss function, respectively. (b)(d) RMSE scores of the throttle, brake, and torque for LSTM-5, Conv-LSTM, and Conv network, respectively, using mse loss function. }
\label{analysis-1}
\end{figure*}

\begin{figure*}[b]
\begin{subfigure}{0.5\textwidth}
\includegraphics[width=7cm]{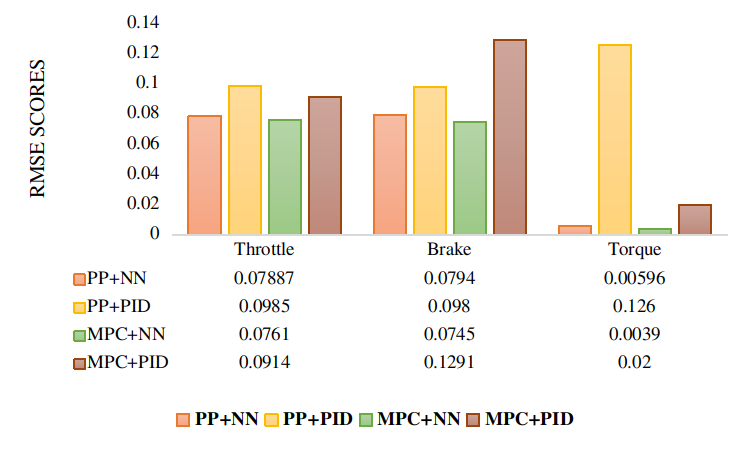} 
\caption{}
\label{LSTM-5-smooth-L1-RMSE}
\end{subfigure}
\quad
\begin{subfigure}{0.5\textwidth}
\includegraphics[width=7cm]{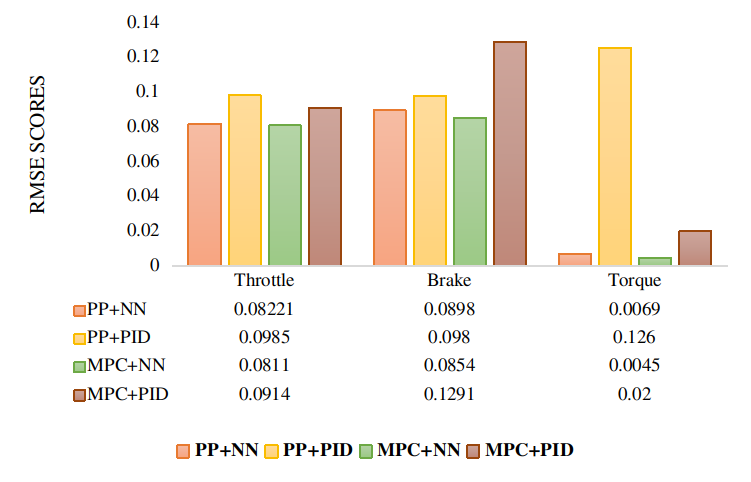}
\caption{}
\label{LSTM-5-mse-RMSE}
\end{subfigure}
\quad
\begin{subfigure}{0.5\textwidth}
\includegraphics[width=7cm]{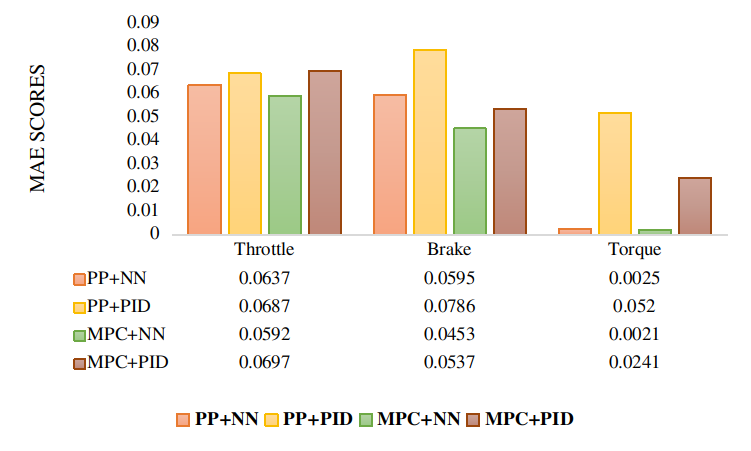} 
\caption{}
\label{HD-smooth-L1-RMSE}
\end{subfigure}
\quad
\begin{subfigure}{0.5\textwidth}
\includegraphics[width=7cm]{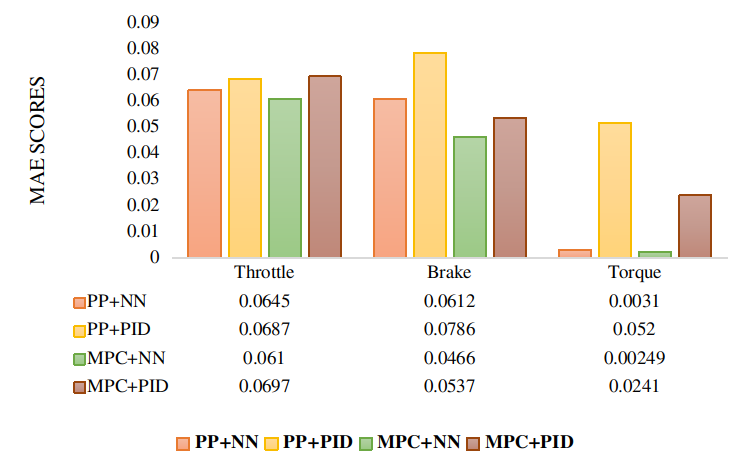} 
\caption{}
\label{LSTM-5-mse-MAE}
\end{subfigure}

\caption{Individual throttle, brake, and torque RMSE and MAE scores comparison between LSTM-5 (proposed method) and PID along with path tracking algorithms (PP and MPC). (a),(c) RMSE and MAE scores for smooth-L1 loss function. (b),(d) RMSE and MAE scores for the mse loss function.}
\label{analysis-2}
\end{figure*}

\begin{figure*}[t]

\begin{subfigure}{0.5\textwidth}
\includegraphics[width=7cm]{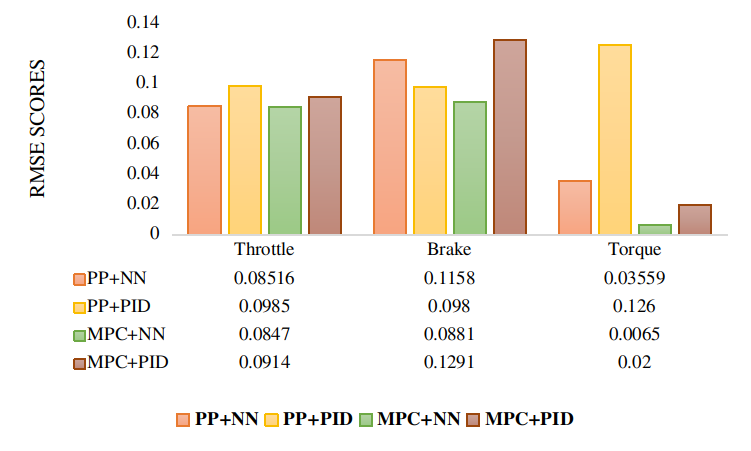} 
\caption{Conv-LSTM-smooth-L1-RMSE}
\label{Conv-LSTM-smooth-L1-RMSE}
\end{subfigure}
\quad
\begin{subfigure}{0.5\textwidth}
\includegraphics[width=7cm]{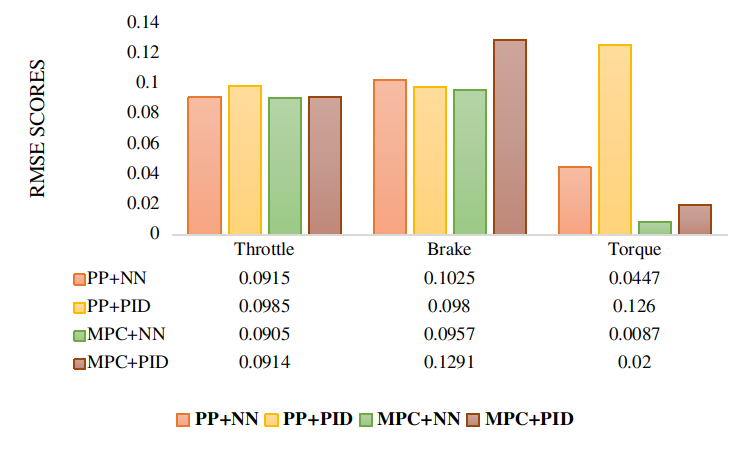}
\caption{Conv-LSTM-mse-RMSE}
\label{Conv-LSTM-mse-RMSE}
\end{subfigure}
\quad
\begin{subfigure}{0.5\textwidth}
\includegraphics[width=7cm]{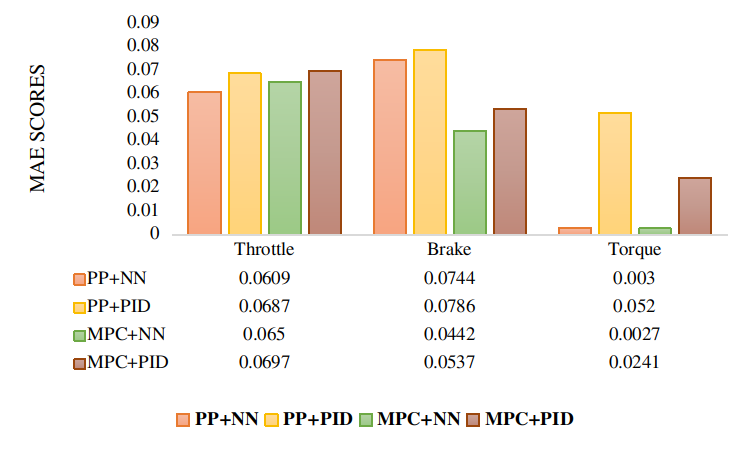} 
\caption{Conv-LSTM-smooth-L1-MAE}
\label{Conv-LSTM-smooth-L1-MAE}
\end{subfigure}
\quad
\begin{subfigure}{0.5\textwidth}
\includegraphics[width=7cm]{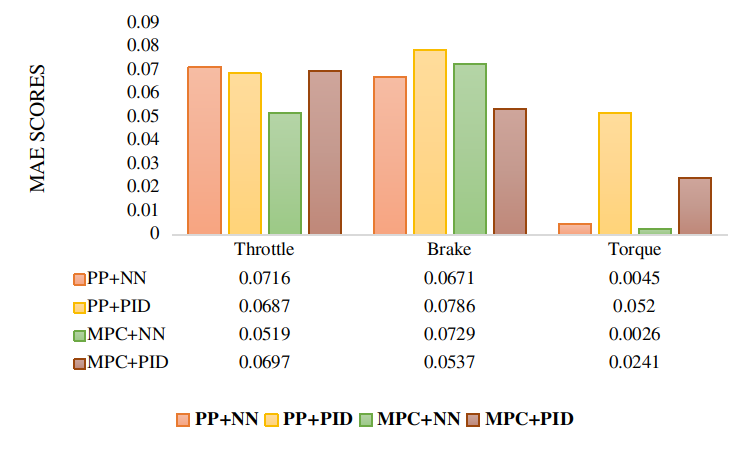} 
\caption{Conv-LSTM-mse-MAE}
\label{Conv-LSTM-mse-MAE}
\end{subfigure}

\caption{Individual throttle, brake, and torque RMSE and MAE scores comparison between Conv-LSTM and PID along with path tracking algorithms (PP and MPC). (a),(c) RMSE and MAE scores for smooth-L1 loss function. (b),(d) RMSE and MAE scores for mse loss function.}
\label{analysis-3}
\end{figure*}

\begin{figure*}[b]
\begin{subfigure}{0.5\textwidth}
\includegraphics[width=0.9\linewidth, height=5cm]{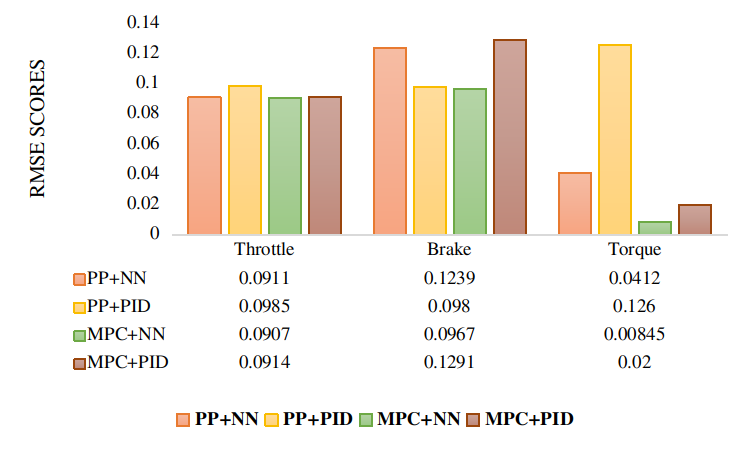} 
\caption{}
\label{Conv-smooth-L1-RMSE}
\end{subfigure}
\quad
\begin{subfigure}{0.5\textwidth}
\includegraphics[width=0.9\linewidth, height=5cm]{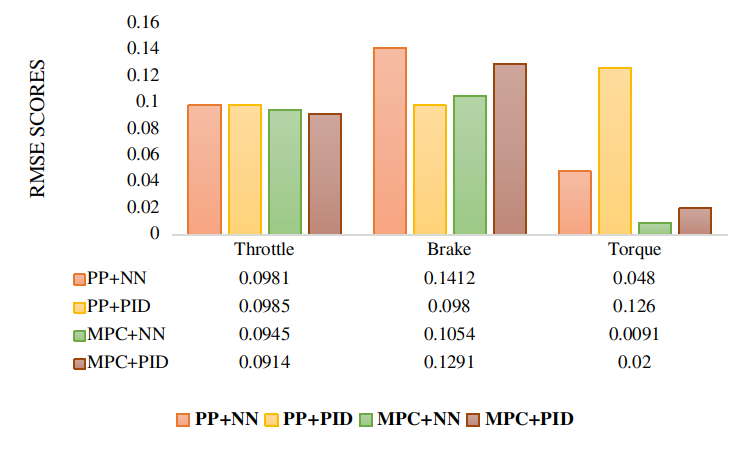}
\caption{}
\label{Conv-mse-RMSE}
\end{subfigure}
\quad
\begin{subfigure}{0.5\textwidth}
\includegraphics[width=0.9\linewidth, height=5cm]{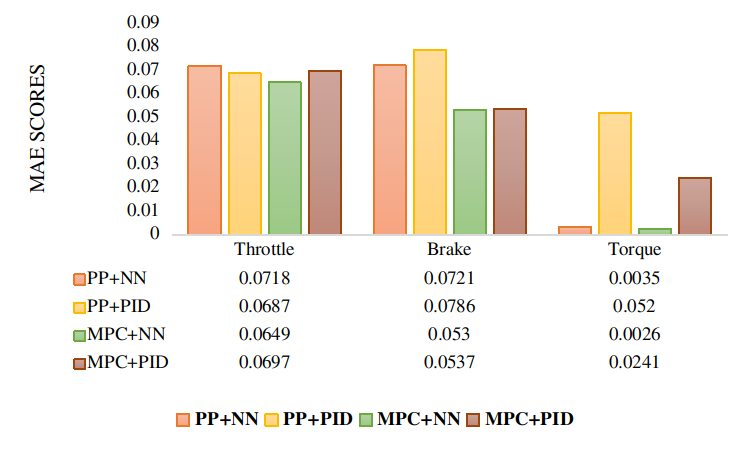} 
\caption{}
\label{Conv-smooth-L1-MAE}
\end{subfigure}
\quad
\begin{subfigure}{0.5\textwidth}
\includegraphics[width=0.9\linewidth, height=5cm]{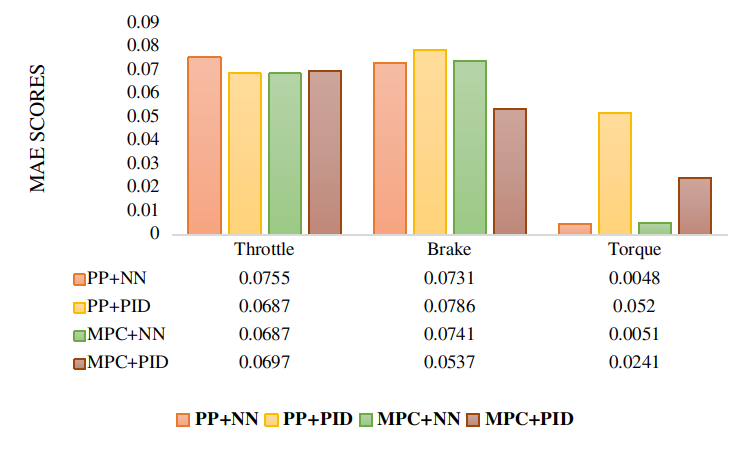} 
\caption{}
\label{Conv-mse-MAE}
\end{subfigure}

\caption{Individual throttle, brake, and torque RMSE and MAE scores comparison between Conv-LSTM and PID along with path tracking algorithms (PP and MPC). (a),(c) RMSE and MAE scores for smooth-L1 loss function. (b),(d) RMSE and MAE scores for mse loss function.}
\label{analysis-4}
\end{figure*}

 
Both of the network configurations are evaluated on test data as well as on the Udacity dataset. Table. \ref{table-1}-Table-\ref{table-3} show the RMSE and MAE scores of these configurations. Fig. \ref{analysis-1} shows the evaluation results of configurations when applied to the human driving test data. The individual analysis of throttle, brake, and torque for both configurations with PID and the path tracking algorithm on collected test data are shown in Fig. \ref{analysis-3} and Fig. \ref{analysis-4} respectively. \blfootnote{Fig. \ref{analysis-1}, Fig. \ref{analysis-2}, Fig. \ref{analysis-3}, and Fig. \ref{analysis-4} are shown at the end of manuscript.}

\begin{table}[h]
\centering
\caption{Mean RMSE and MAE score for Speed and Steering angle prediction. }
\label{table-4}
\resizebox{4.5cm}{!}{%
\begin{tabular}{@{}cll@{}}
\toprule
\multicolumn{1}{l}{}                 & \multicolumn{2}{c}{Human Driving}                  \\ \midrule
\multicolumn{1}{c|}{Dataset}         & \multicolumn{1}{c}{RMSE} & \multicolumn{1}{c}{MAE} \\ \midrule
\multicolumn{1}{c|}{Ours}            & \textbf{0.03045}          & \textbf{0.021}          \\
\multicolumn{1}{c|}{Udacity} & 0.03325                   & 0.026                   \\ \bottomrule
\end{tabular}%
}
\end{table}

\subsection{End-to-End Prediction of Steering Angle and Speed}
In this section, the experimental details of the proposed method for predicting the steering angle and speed are discussed using image data.
\par 
The proposed network is trained for both datasets and include weather conditions for instance (sunny, overcast, shadows). The training data in both datasets consists of the pair of images and speed and steering angle values. Before training the network, the input image is downsampled to $400 \times 400$ dimension for efficiency purposes. The mean and standard deviation normalization is used for the input images. Also, data augmentation is also applied to input data which are as follows:
\begin{enumerate}
    \item Randomly change the brightness by a factor of $0.2$.
    \item Randomly change the rotation of the input image by $20$ degrees and shear angle randomly by $0.15$ degrees.
    \item Randomly flip the input image horizontally 
\end{enumerate}
In training, Adam optimizer is used with a learning rate of $0.001$ with a decay rate of $0.0001$. The model is trained for $300$ epochs having the batch size of $12$. The mean squared error (MSE) is used as a loss criterion for both speed and steering angle, and the overall model loss is the summation of both speed loss and steering angle loss.
\par 
For the quantitative analysis, RMSE and MAE scores are calculated on test data of the collected dataset. The proposed network for predicting the steering angle and speed are also evaluated on the Udacity dataset. Table- \ref{table-4} shows the quantitative results on our data as well as on the Udacity dataset. The proposed method is also compared with other state-of-the-art algorithms. Table-\ref{table-5} represents the RMSE score of the proposed method, along with other state-of-the-art algorithms. Fig. \ref{qualitative-ours} and Fig. \ref{qualitative-udacity} shows the quantitative results of speed and steering angle prediction on our dataset as well as on the Udacity dataset, respectively. 

\begin{table}[]
\centering
\caption{Comparison of RMSE scores of steering angle and speed prediction. * indicates the MSE scores are given for these algorithms which are converted to RMSE for comparison with our method.\dag shows the models tested on Udacity dataset.}
\label{table-5}
\resizebox{9cm}{!}{%
\begin{tabular}{@{}cccl@{}}
\toprule
Model                            & RMSE (Steering Angle) & \multicolumn{2}{c}{RMSE (Speed)} \\ \midrule
Zero$^\dag$ \cite{chi2017deep}                     & 0.2077                & \multicolumn{2}{c}{-}               \\
Mean$^\dag$ \cite{chi2017deep}                     & 0.2098                & \multicolumn{2}{c}{-}               \\
AlexNet$^\dag$ \cite{chi2017deep}                  & 0.1299                & \multicolumn{2}{c}{-}               \\
PilotNet$^\dag$ \cite{chi2017deep}                 & 0.1604                & \multicolumn{2}{c}{-}               \\
VGG-16$^\dag$ \cite{chi2017deep}                   & 0.0948                & \multicolumn{2}{c}{-}               \\
(ST-Conv + ConvLSTM +LSTM)$^\dag$ \cite{chi2017deep} & 0.0948                & \multicolumn{2}{c}{-}               \\
\cite{zhao2017speed}                          & 0.0677                & \multicolumn{2}{c}{0.99206}         \\
\cite{hecker2018end}*                         & 0.3660                & \multicolumn{2}{c}{0.05477}         \\
\cite{xu2017end}*                         & 0.4012                & \multicolumn{2}{c}{0.0812}          \\
Ours (collected dataset)         & \textbf{0.0249}       & \multicolumn{2}{c}{\textbf{0.036}}  \\
Ours (Udacity dataset)           & \textbf{0.0298}       & \multicolumn{2}{c}{\textbf{0.0387}} \\ \bottomrule
\end{tabular}%
}
\end{table}

\begin{table}[b]
\centering
\caption{RMSE and MAE scores of unified network with smooth-L1 and mse loss function on collected and Udacity datasets.}
\label{table-6}
\resizebox{7cm}{!}{
\begin{tabular}{@{}c|cccc@{}}
\toprule
Loss function & \multicolumn{2}{c|}{Smooth L1}                       & \multicolumn{2}{c}{MSE}         \\ \midrule
Dataset       & \multicolumn{1}{c|}{RMSE} & \multicolumn{1}{c|}{MAE} & \multicolumn{1}{c|}{RMSE} & MAE \\ \midrule
Collected & 0.05136 & 0.03097 & 0.0567  & 0.035 \\
Udacity   & 0.05741 & 0.0336  & 0.06874 & 0.043 \\ \bottomrule
\end{tabular}
}
\end{table}

\subsection{ End-to-End prediction of speed and Steering angle with $N^2C$}
A unified framework that includes the prediction of steering angle and speed along with the neural network controller, $N^2C$, has experimented in this section. In this framework, the predicted speed and steering angle is passed to the $N^2C$ to predict the throttle, brake, and torque. For the evaluation, this architecture is tested on both datasets. Table-\ref{table-6} shows the RMSE and MAE scores. The effect of the loss function in $N^2C$ is analyzed by computing the metric scores for both loss function, which are used in the design of $N^2C$.

% \usepackage{booktabs}
% \usepackage{graphicx}

% \usepackage{booktabs}
% \usepackage{graphicx}

% \usepackage{booktabs}
% \usepackage{graphicx}
% \begin{table}[]
% \centering
% \caption{}
% \label{table-7}
% \resizebox{5cm}{!}{%
% \begin{tabular}{@{}lcl@{}}
% \toprule
% \multicolumn{3}{c}{MSE loss function}                            \\ \midrule
% \multicolumn{1}{l|}{Dataset} & RMSE    & \multicolumn{1}{c}{MAE} \\ \midrule
% Ours                         & 0.0567  & 0.035                   \\
% Udacity                      & 0.06874 & 0.043                   \\ \bottomrule
% \end{tabular}%
% }
% \end{table}

% \usepackage{booktabs}
% \usepackage{graphicx}

\section{Conclusion}
In this paper, we proposed a controller ($N^2C$) for an autonomous vehicle constituted with an artificial neural network. The proposed controller ($N^2C$) is inspired by behavioral cloning, and prospects as a replacement of the classical control module in the autonomous vehicles. The designed neural network predicts throttle, brake, and torque by taking steering angle and speed as inputs. The experimental evaluation of the controller exhibits its usability in lieu of conventional classical controller when experimented with the two path tracking algorithms: pure pursuit and model predictive control based path follower. The evaluation results show better RMSE and MAE scores for $N^2C$ in contrast to the conventional PID controller. The second part presented in this study is focused on the design of an end-to-end ANN that predicts steering angle and speed from images. The experimental results show the better metric scores of end-to-end ANN in predicting speed and steering angle on the collected and Udacity datasets compared to state-of-the-art methods. Further, as a supplement, the $N^2C$ is also tested with an end-to-end steering angle and speed prediction ANN in a unified framework. The simulation results, in terms of better metric scores, show the efficacy that the unified framework provides a competitive replacement for the conventional pipeline of motion planning and control. In the proposed work, the $N^2C$ predicts the throttle, brake and torques by taking the average wheel speed and steering angle as input. The possible limitations of the proposed work is to include other source of obtaining the speed. The fusion of speed from GNSS (also gives the average speed) and wheel speed in order to enhance the current experimentation results is one possible future work. In addition, the incorporation of dynamics of vehicles by along the lines of proposed work is another future aspect of the proposed work. The future work also includes the curriculum learning of the proposed controller with the inclusion of more human behaviours in order to simulate the controller act like human driving rather than a computerized controller. In addition, the fusion of observations from human steering using electromyography sensors with CAN bus data for the controller design is also a future prospect of this work.

\section*{Acknowledgments}
This work was partly supported by the ICT R$\&$D program of MSIP/IITP (2014-0-00077, Development of global multi-target tracking and event prediction techniques based on real-time large-scale video analysis), Institute of Information and Communications Technology Planning and Evaluation (IITP) grant funded by the Korea governemnt (MSIT)(No.2019-0-01842, Artificial Intelligence Graduate School Program (GIST)), Ministry of Culture, Sports and Tourism (MCST), and Korea Creative Content Agency (KOCCA) in the Culture Technology (CT) Research \& Development (R2020070004) Program 2020.

% Moreover, the robustness of the neural-based controller is tested with the end-to-end ANN steering angle and speed predicting module.
 
\ifCLASSOPTIONcaptionsoff
  \newpage
\fi

\bibliographystyle{IEEEtran}
\bibliography{ref}

\begin{thebibliography}{10}
\providecommand{\url}[1]{#1}
\csname url@samestyle\endcsname
\providecommand{\newblock}{\relax}
\providecommand{\bibinfo}[2]{#2}
\providecommand{\BIBentrySTDinterwordspacing}{\spaceskip=0pt\relax}
\providecommand{\BIBentryALTinterwordstretchfactor}{4}
\providecommand{\BIBentryALTinterwordspacing}{\spaceskip=\fontdimen2\font plus
\BIBentryALTinterwordstretchfactor\fontdimen3\font minus
  \fontdimen4\font\relax}
\providecommand{\BIBforeignlanguage}[2]{{%
\expandafter\ifx\csname l@#1\endcsname\relax
\typeout{** WARNING: IEEEtran.bst: No hyphenation pattern has been}%
\typeout{** loaded for the language `#1'. Using the pattern for}%
\typeout{** the default language instead.}%
\else
\language=\csname l@#1\endcsname
\fi
#2}}
\providecommand{\BIBdecl}{\relax}
\BIBdecl

\bibitem{kiencke1986automotive}
U.~Kiencke, S.~Dais, and M.~Litschel, ``Automotive serial controller area
  network,'' \emph{SAE transactions}, pp. 823--828, 1986.

\bibitem{munir2019localization}
F.~Munir, S.~Azam, A.~M. Sheri, Y.~Ko, and M.~Jeon, ``Where am i: Localization
  and 3d maps for autonomous vehicles.'' in \emph{VEHITS}, 2019, pp. 452--457.

\bibitem{azam2020dynamic}
S.~Azam, F.~Munir, and M.~Jeon, ``Dynamic control system design for autonomous
  car.'' in \emph{VEHITS}, 2020, pp. 456--463.

\bibitem{farhadi2018yolov3}
A.~Farhadi and J.~Redmon, ``Yolov3: An incremental improvement,''
  \emph{Computer Vision and Pattern Recognition, cite as}, 2018.

\bibitem{liu2016ssd}
W.~Liu, D.~Anguelov, D.~Erhan, C.~Szegedy, S.~Reed, C.-Y. Fu, and A.~C. Berg,
  ``Ssd: Single shot multibox detector,'' in \emph{European conference on
  computer vision}.\hskip 1em plus 0.5em minus 0.4em\relax Springer, 2016, pp.
  21--37.

\bibitem{badrinarayanan2017segnet}
V.~Badrinarayanan, A.~Kendall, and R.~Cipolla, ``Segnet: A deep convolutional
  encoder-decoder architecture for image segmentation,'' \emph{IEEE
  transactions on pattern analysis and machine intelligence}, vol.~39, no.~12,
  pp. 2481--2495, 2017.

\bibitem{lang2019pointpillars}
A.~H. Lang, S.~Vora, H.~Caesar, L.~Zhou, J.~Yang, and O.~Beijbom,
  ``Pointpillars: Fast encoders for object detection from point clouds,'' in
  \emph{Proceedings of the IEEE Conference on Computer Vision and Pattern
  Recognition}, 2019, pp. 12\,697--12\,705.

\bibitem{perez2011cascade}
J.~P{\'e}rez, V.~Milan{\'e}s, and E.~Onieva, ``Cascade architecture for lateral
  control in autonomous vehicles,'' \emph{IEEE Transactions on Intelligent
  Transportation Systems}, vol.~12, no.~1, pp. 73--82, 2011.

\bibitem{cristi1990adaptive}
R.~Cristi, F.~A. Papoulias, and A.~J. Healey, ``Adaptive sliding mode control
  of autonomous underwater vehicles in the dive plane,'' \emph{IEEE journal of
  Oceanic Engineering}, vol.~15, no.~3, pp. 152--160, 1990.

\bibitem{borrelli2005mpc}
F.~Borrelli, P.~Falcone, T.~Keviczky, J.~Asgari, and D.~Hrovat, ``Mpc-based
  approach to active steering for autonomous vehicle systems,''
  \emph{International journal of vehicle autonomous systems}, vol.~3, no. 2-4,
  pp. 265--291, 2005.

\bibitem{torabi2018behavioral}
F.~Torabi, G.~Warnell, and P.~Stone, ``Behavioral cloning from observation,''
  \emph{arXiv preprint arXiv:1805.01954}, 2018.

\bibitem{zhang2018behavioral}
X.~Zhang, M.~Chen, and X.~Zhan, ``Behavioral cloning for driverless cars using
  transfer learning,'' in \emph{2018 IEEE/ION Position, Location and Navigation
  Symposium (PLANS)}.\hskip 1em plus 0.5em minus 0.4em\relax IEEE, 2018, pp.
  1069--1073.

\bibitem{qiu2019learning}
T.~Qiu and Z.~Huang, ``Learning a steering decision policy for end-to-end
  control of autonomous vehicle,'' in \emph{2019 5th International Conference
  on Control, Automation and Robotics (ICCAR)}.\hskip 1em plus 0.5em minus
  0.4em\relax IEEE, 2019, pp. 347--351.

\bibitem{munir2018autonomous}
F.~Munir, S.~Azam, M.~I. Hussain, A.~M. Sheri, and M.~Jeon, ``Autonomous
  vehicle: The architecture aspect of self driving car,'' in \emph{Proceedings
  of the 2018 International Conference on Sensors, Signal and Image
  Processing}, 2018, pp. 1--5.

\bibitem{kritayakirana2012autonomous}
K.~Kritayakirana and J.~C. Gerdes, ``Autonomous vehicle control at the limits
  of handling,'' \emph{International Journal of Vehicle Autonomous Systems},
  vol.~10, no.~4, pp. 271--296, 2012.

\bibitem{khodayari2010historical}
A.~Khodayari, A.~Ghaffari, S.~Ameli, and J.~Flahatgar, ``A historical review on
  lateral and longitudinal control of autonomous vehicle motions,'' in
  \emph{2010 International Conference on Mechanical and Electrical
  Technology}.\hskip 1em plus 0.5em minus 0.4em\relax IEEE, 2010, pp. 421--429.

\bibitem{kodagoda2002fuzzy}
K.~Kodagoda, W.~S. Wijesoma, and E.~K. Teoh, ``Fuzzy speed and steering control
  of an agv,'' \emph{IEEE Transactions on control systems technology}, vol.~10,
  no.~1, pp. 112--120, 2002.

\bibitem{wang2019speed}
J.~F. Wang and H.~Zhao, ``Speed control of tracked vehicle autonomous driving
  system using fuzzy self-tuning pid,'' in \emph{2019 4th International
  Conference on Mechanical, Control and Computer Engineering (ICMCCE)}.\hskip
  1em plus 0.5em minus 0.4em\relax IEEE, 2019, pp. 305--3053.

\bibitem{han2019design}
X.~Han, X.~Zhang, Y.~Du, and G.~Cheng, ``Design of autonomous vehicle
  controller based on bp-pid,'' in \emph{IOP Conference Series: Earth and
  Environmental Science}, vol. 234, no.~1.\hskip 1em plus 0.5em minus
  0.4em\relax IOP Publishing, 2019, p. 012097.

\bibitem{wang2020adaptive}
J.~Wang, Y.~Zhu, R.~Qi, X.~Zheng, and W.~Li, ``Adaptive pid control of
  multi-dof industrial robot based on neural network,'' \emph{Journal of
  Ambient Intelligence and Humanized Computing}, pp. 1--12, 2020.

\bibitem{falcone2008mpc}
P.~Falcone, H.~Eric~Tseng, F.~Borrelli, J.~Asgari, and D.~Hrovat, ``Mpc-based
  yaw and lateral stabilisation via active front steering and braking,''
  \emph{Vehicle System Dynamics}, vol.~46, no.~S1, pp. 611--628, 2008.

\bibitem{hu2020lane}
J.~Hu, S.~Xiong, J.~Zha, and C.~Fu, ``Lane detection and trajectory tracking
  control of autonomous vehicle based on model predictive control,''
  \emph{International journal of automotive technology}, vol.~21, no.~2, pp.
  285--295, 2020.

\bibitem{tosolin2020development}
G.~Tosolin, J.~Cartr{\'o}, and V.~Sharma, ``Development of model predictive
  motion planning and control for autonomous vehicles,'' in \emph{10th
  International Munich Chassis Symposium 2019}.\hskip 1em plus 0.5em minus
  0.4em\relax Springer, 2020, pp. 323--340.

\bibitem{alcala2018autonomous}
E.~Alcala, V.~Puig, J.~Quevedo, T.~Escobet, and R.~Comasolivas, ``Autonomous
  vehicle control using a kinematic lyapunov-based technique with lqr-lmi
  tuning,'' \emph{Control engineering practice}, vol.~73, pp. 1--12, 2018.

\bibitem{urmson2006robust}
C.~Urmson, C.~Ragusa, D.~Ray, J.~Anhalt, D.~Bartz, T.~Galatali, A.~Gutierrez,
  J.~Johnston, S.~Harbaugh, H.~“Yu”~Kato \emph{et~al.}, ``A robust approach
  to high-speed navigation for unrehearsed desert terrain,'' \emph{Journal of
  Field Robotics}, vol.~23, no.~8, pp. 467--508, 2006.

\bibitem{thrun2006stanley}
S.~Thrun, M.~Montemerlo, H.~Dahlkamp, D.~Stavens, A.~Aron, J.~Diebel, P.~Fong,
  J.~Gale, M.~Halpenny, G.~Hoffmann \emph{et~al.}, ``Stanley: The robot that
  won the darpa grand challenge,'' \emph{Journal of field Robotics}, vol.~23,
  no.~9, pp. 661--692, 2006.

\bibitem{urmson2008autonomous}
C.~Urmson, J.~Anhalt, D.~Bagnell, C.~Baker, R.~Bittner, M.~Clark, J.~Dolan,
  D.~Duggins, T.~Galatali, C.~Geyer \emph{et~al.}, ``Autonomous driving in
  urban environments: Boss and the urban challenge,'' \emph{Journal of Field
  Robotics}, vol.~25, no.~8, pp. 425--466, 2008.

\bibitem{huang2018apolloscape}
X.~Huang, X.~Cheng, Q.~Geng, B.~Cao, D.~Zhou, P.~Wang, Y.~Lin, and R.~Yang,
  ``The apolloscape dataset for autonomous driving,'' in \emph{Proceedings of
  the IEEE Conference on Computer Vision and Pattern Recognition Workshops},
  2018, pp. 954--960.

\bibitem{pomerleau1989autonomous}
D.~Pomerleau, ``An autonomous land vehicle in a neural network. nips 1,'' 1989.

\bibitem{bojarski2016end}
M.~Bojarski, D.~Del~Testa, D.~Dworakowski, B.~Firner, B.~Flepp, P.~Goyal, L.~D.
  Jackel, M.~Monfort, U.~Muller, J.~Zhang \emph{et~al.}, ``End to end learning
  for self-driving cars,'' \emph{arXiv preprint arXiv:1604.07316}, 2016.

\bibitem{du2019self}
S.~Du, H.~Guo, and A.~Simpson, ``Self-driving car steering angle prediction
  based on image recognition,'' \emph{arXiv preprint arXiv:1912.05440}, 2019.

\bibitem{chi2017deep}
L.~Chi and Y.~Mu, ``Deep steering: Learning end-to-end driving model from
  spatial and temporal visual cues,'' \emph{arXiv preprint arXiv:1708.03798},
  2017.

\bibitem{chowdhuri2019multinet}
S.~Chowdhuri, T.~Pankaj, and K.~Zipser, ``Multinet: Multi-modal multi-task
  learning for autonomous driving,'' in \emph{2019 IEEE Winter Conference on
  Applications of Computer Vision (WACV)}.\hskip 1em plus 0.5em minus
  0.4em\relax IEEE, 2019, pp. 1496--1504.

\bibitem{yang2018end}
Z.~Yang, Y.~Zhang, J.~Yu, J.~Cai, and J.~Luo, ``End-to-end multi-modal
  multi-task vehicle control for self-driving cars with visual perceptions,''
  in \emph{2018 24th International Conference on Pattern Recognition
  (ICPR)}.\hskip 1em plus 0.5em minus 0.4em\relax IEEE, 2018, pp. 2289--2294.

\bibitem{kim2017interpretable}
J.~Kim and J.~Canny, ``Interpretable learning for self-driving cars by
  visualizing causal attention,'' in \emph{Proceedings of the IEEE
  international conference on computer vision}, 2017, pp. 2942--2950.

\bibitem{xu2017end}
H.~Xu, Y.~Gao, F.~Yu, and T.~Darrell, ``End-to-end learning of driving models
  from large-scale video datasets,'' in \emph{Proceedings of the IEEE
  conference on computer vision and pattern recognition}, 2017, pp. 2174--2182.

\bibitem{hou2019learning}
Y.~Hou, Z.~Ma, C.~Liu, and C.~C. Loy, ``Learning to steer by mimicking features
  from heterogeneous auxiliary networks,'' in \emph{Proceedings of the AAAI
  Conference on Artificial Intelligence}, vol.~33, 2019, pp. 8433--8440.

\bibitem{zhao2017pyramid}
H.~Zhao, J.~Shi, X.~Qi, X.~Wang, and J.~Jia, ``Pyramid scene parsing network,''
  in \emph{Proceedings of the IEEE conference on computer vision and pattern
  recognition}, 2017, pp. 2881--2890.

\bibitem{ilg2017flownet}
E.~Ilg, N.~Mayer, T.~Saikia, M.~Keuper, A.~Dosovitskiy, and T.~Brox, ``Flownet
  2.0: Evolution of optical flow estimation with deep networks,'' in
  \emph{Proceedings of the IEEE conference on computer vision and pattern
  recognition}, 2017, pp. 2462--2470.

\bibitem{gers2002learning}
F.~A. Gers, N.~N. Schraudolph, and J.~Schmidhuber, ``Learning precise timing
  with lstm recurrent networks,'' \emph{Journal of machine learning research},
  vol.~3, no. Aug, pp. 115--143, 2002.

\bibitem{zhao2017speed}
C.~Zhao, J.~Gong, C.~Lu, G.~Xiong, and W.~Mei, ``Speed and steering angle
  prediction for intelligent vehicles based on deep belief network,'' in
  \emph{2017 IEEE 20th International Conference on Intelligent Transportation
  Systems (ITSC)}.\hskip 1em plus 0.5em minus 0.4em\relax IEEE, 2017, pp.
  301--306.

\bibitem{hecker2018end}
S.~Hecker, D.~Dai, and L.~Van~Gool, ``End-to-end learning of driving models
  with surround-view cameras and route planners,'' in \emph{Proceedings of the
  european conference on computer vision (eccv)}, 2018, pp. 435--453.

\end{thebibliography}
\begin{IEEEbiography}[{\includegraphics[width=1in,height=1.25in,clip,keepaspectratio]{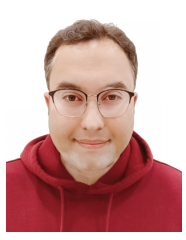}}]{Shoaib Azam} received a B.S. degree in Engineering Sciences from Ghulam Ishaq Khan Institute of Science and Technology, Pakistan, in 2010, and an M.S. degree in Robotics and Intelligent Machine Engineering from National
University of Science and Technology, Pakistan, in 2015. He is currently
pursuing a Ph.D. with the Department of Electrical Engineering and
Computer Science, Gwangju Institute of Science and Technology, Gwangju,
South Korea. His current research interests include deep learning and autonomous driving.
\end{IEEEbiography}

\begin{IEEEbiography}[{\includegraphics[width=1in,height=1.25in,clip,keepaspectratio]{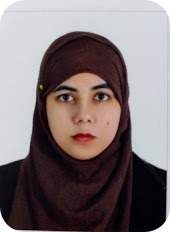}}]{Farzeen Munir} received the B.S. degree in Electrical Engineering from Pakistan Institute of Engineering and Applied Sciences, Pakistan in 2013,
and MS degree in System Engineering from Pakistan Institute of Engineering
and Applied Sciences, Pakistan in 2015. Now she is pursing her PhD
degree at Gwangju Institute of Science and Technology, Korea in Electrical
Engineering and Computer Science. Her current research interest include,
machine Learning, deep neural network, autonomous driving and computer vision. 
\end{IEEEbiography}

\begin{IEEEbiography}[{\includegraphics[width=1in,height=1.25in,clip,keepaspectratio]{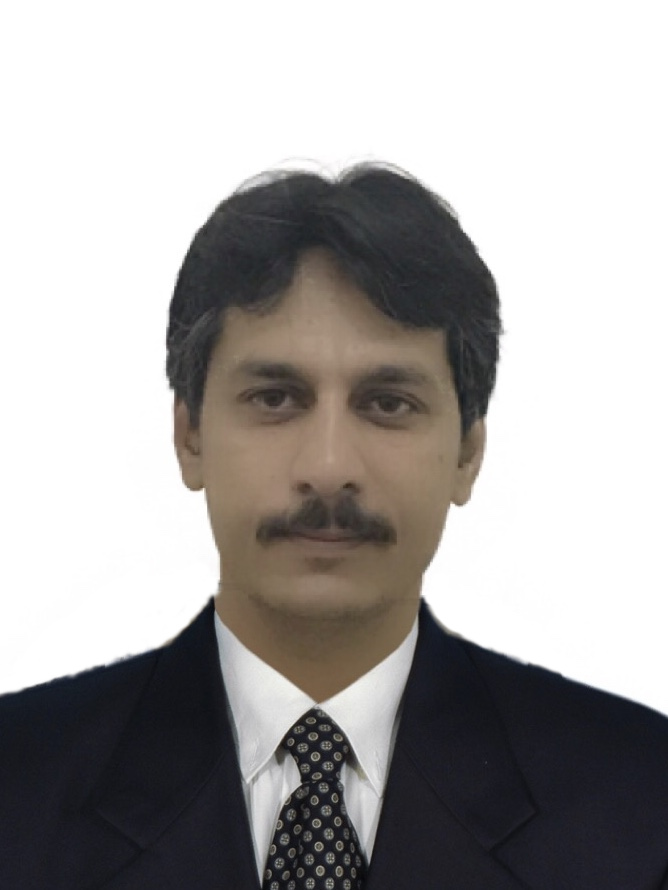}}]{Aasim Rafique} received his M.Sc. degree in computer science from Quaid-e-Azam University, Islamabad, Pakistan. He then received his M.Sc. degree in computer science
from Lahore University of Management and Sciences, Lahore,
Pakistan in 2008. He received his Ph.D. degree from School of
Electrical Engineering and Computer Sciences, GIST, Gwangju,
Republic of Korea, in 2018. He is now working as Assistant Professor at School of Electrical Engineering and Computer Science, National University of Sciences and Technology (NUST), Pakistan. His research interests are artificial neural networks, their application
in machine learning and computer vision.
\end{IEEEbiography}

\begin{IEEEbiography}[{\includegraphics[width=1in,height=1.25in,clip,keepaspectratio]{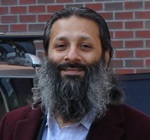}}]{Ahmad Muqeem Sheri} received the B.E. degree in computer software engineering from the National University of Science and Technology, Rawalpindi,
Pakistan, and the M.S. degree in computer sciences from the Lahore University
of Management and Sciences, Lahore, Pakistan, in 2008. He received his
PhD degree in Computer Science from the School of Information and
Communication, Gwangju Institute of Science and Technology, Gwangju,
Korea. Currently he is working as an Assistant Professor in Department of Computer Software Engineering, National University of Sciences and Technology (NUST), Pakistan.
\end{IEEEbiography}

\begin{IEEEbiography}[{\includegraphics[width=1in,height=1.25in,clip,keepaspectratio]{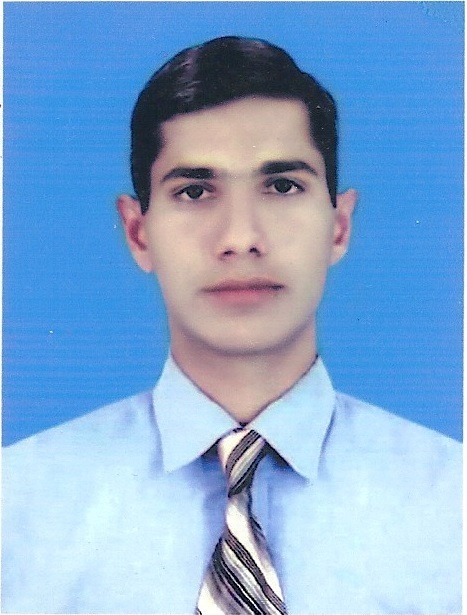}}]{Muhammad Ishfaq Hussain} received the B.Sc, degree from University of The Punjab Lahore, Pakistan in 2008, and M.S. degree in Computer Software Engineering from National
University of Science and Technology, Pakistan in 2016. He is currently
pursuing the Ph.D. degree with the Department of Electrical Engineering and
Computer Science, Gwangju Institute of Science and Technology, Gwangju,
South Korea. His current research interests include artificial intelligence and
machine learning, sensor fusion, robotics, and autonomous driving.
\end{IEEEbiography}

\begin{IEEEbiography}[{\includegraphics[width=1in,height=1.25in,clip,keepaspectratio]{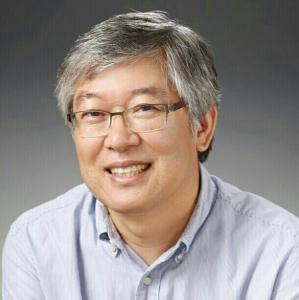}}]{Moongu Jeon} received the B.S. degree in architectural engineering from Korea University, Seoul, South Korea, in 1988, and the M.S. and Ph.D.
degrees in computer science and scientific computation from the University
of Minnesota, Minneapolis, MN, USA, in 1999 and 2001, respectively. As
a Post-Graduate Researcher, he worked on optimal control problems at the
University of California at Santa Barbara, Santa Barbara, CA, USA, from
2001 to 2003, and then moved to the National Research Council of Canada,
where he worked on the sparse representation of high-dimensional data and
the image processing until 2005. In 2005, he joined the Gwangju Institute of
Science and Technology, Gwangju, South Korea, where he is currently a Full
Professor with the School of Electrical Engineering and Computer Science.
His current research interests are in machine learning, computer vision, and
artificial intelligence.
\end{IEEEbiography}

% \end{IEEEbiography}


% % insert where needed to balance the two columns on the last page with
% %\newpage



%\vfill


\end{document}